\documentclass{vldb}
\usepackage{balance}  
\usepackage{multirow}
\usepackage{graphicx} 
\usepackage[hyphens]{url}      
\usepackage{amsmath}
\usepackage{color}
\usepackage{cancel}
\usepackage{listings}
\usepackage[normalem]{ulem}
\usepackage{graphicx}
\usepackage{subcaption}
\usepackage{booktabs}
\usepackage{graphicx}
\usepackage[ruled,vlined,algonl,boxed]{algorithm2e}
\usepackage{algorithmic}
\usepackage{wrapfig}
\usepackage{enumitem}
\usepackage{xspace}
\usepackage{xcolor}
\usepackage{colortbl}
\usepackage[T1]{fontenc}
\usepackage{fancyvrb}
\usepackage[bottom]{footmisc}

\usepackage[nocompress]{cite}
\usepackage{microtype}
\usepackage[section]{placeins}

\makeatletter
\def\url@leostyle{
  \@ifundefined{selectfont}{\def\UrlFont{\sf}}{\def\UrlFont{\small\bf\ttfamily}}}
\makeatother
\makeatletter
\def\@copyrightspace{\relax}
\makeatother

\def\pprw{8.5in}
\def\pprh{11in}

\newcounter{prob}
\newtheorem{problem}[prob]{Problem}

\usepackage{soul,xcolor}

\DeclareMathOperator*{\argmin}{arg\,min}
\DeclareMathOperator*{\argmax}{arg\,max}
\DeclareMathOperator*{\topk}{topk}

\usepackage[pdftex]{hyperref}
\hypersetup{
  colorlinks=false,
  linkcolor=darkred,
  citecolor=darkgreen,
  urlcolor=darkblue
}

\definecolor{light-gray}{gray}{0.95}
\definecolor{mid-gray}{gray}{0.85}
\definecolor{darkred}{rgb}{0.7,0.25,0.25}
\definecolor{darkgreen}{rgb}{0.15,0.55,0.15}
\definecolor{darkblue}{rgb}{0.1,0.1,0.5}
\definecolor{blue}{rgb}{0.19,0.58,1}
\definecolor{purple}{rgb}{0.43,0.18,.65}

\newcommand{\green}[1]{\textcolor{darkgreen}{#1}}
\newcommand{\purple}[1]{\textcolor{purple}{#1}}
\newcommand{\blue}[1]{\textcolor{blue}{#1}}


\makeatletter
\def\maketag@@@#1{\hbox{\m@th\normalfont\normalsize#1}}
\DeclareRobustCommand*\textsubscript[1]{
          \@textsubscript{\selectfont#1}}
        \def\@textsubscript#1{
          {\m@th\ensuremath{_{\mbox{\fontsize\sf@size\z@#1}}}}}
\makeatother

\newcommand{\stitle}[1]{\vspace{0.5em}\noindent\textbf{#1}}

\newcommand{\sol}{\texttt{TCruise}\xspace}
\newcommand{\sys}{\textsf{Precog}\xspace}
\newcommand{\qp}{\texttt{Precog}\xspace}
\newcommand{\prehoc}{pre-hoc quality control\xspace}
\newcommand{\posthoc}{post-hoc quality control\xspace}
\newcommand{\sep}{Segment-Predict-Explain\xspace}
\newcommand{\isep}{{\it Segment-Predict-Explain}\xspace}

\setstcolor{red} 

\begin{document}

\title{PreCog: Improving Crowdsourced Data Quality \\Before Acquisition}

\numberofauthors{3}
\author{
	\begin{tabular}{ccc}
	Hamed Nilforoshan & Jiannan Wang\large{$^{\diamondsuit}$} & Eugene Wu 
	\end{tabular}
	\and 
	\begin{tabular}{cc}
	Columbia University & Simon Fraser University\large{$^{\diamondsuit}$} \\
	\{hn2284, ew2493\}@columbia.edu & jnwang@sfu.ca
	\end{tabular}
}

\maketitle

\begin{abstract}
  \looseness -1
  Quality control in crowdsourcing systems is crucial.  It is typically done after data collection, often using additional crowdsourced tasks to assess and improve the quality.  These post-hoc methods can easily add cost and latency to the acquisition process---particularly if collecting high-quality data is important.  In this paper, we argue for pre-hoc interface optimizations based on feedback that helps workers improve data quality before it is submitted and is well suited to complement post-hoc techniques.  We propose the \sys system that explicitly supports such interface optimizations for common integrity constraints as well as more ambiguous text acquisition tasks where quality is ill-defined.  We then develop the {\it Segment-Predict-Explain} pattern for detecting low-quality text segments and generating {\it prescriptive} explanations to help the worker improve their text input.  Our unique combination of segmentation and {\it prescriptive} explanation are necessary for \sys to collect $2\times$ more high-quality text data than non-\sys approaches on two real domains.

\end{abstract}

\section{Introduction}
\label{s:intro}

A dominant use case for crowdsourcing is to collect data---labels, opinions, text extraction, ratings---from large groups of workers.
Although crowdsourcing is used to collect labels and simple data for machine learning applications, many popular online communities such as Amazon, AirBnB, Quora, Reddit, and others also rely on collecting and presenting high quality, open-ended content that is crowdsourced from their users.
For example, Amazon crowdsources product reviews by asking customers to rate products and write reviews for them;
rental services (e.g., AirbnB) relies on rental hosts to describe their rental properties in quantitative (e.g., number of bed rooms, wireless) as well as qualitative terms (e.g., textual description).

Quality control for crowdsourcing has been extensively studied~\cite{DBLP:journals/tkde/LiWZF16} and can be modeled in two phases. {\it Pre-hoc} methods improve quality before the data is acquired (submitted); {\it Post-hoc} methods improve quality after data acquisition (i.e., after submission). Most studies focus on \emph{post-hoc} quality control, often using additional crowdsourced tasks to assess and improve the quality. For example, task replication~\cite{DBLP:conf/sigmod/SarmaPW16,ipeirotis2010quality} assigns the same task to multiple workers and aggregates them into a single result; multi-stage workflow design~\cite{bernstein2010soylent,kittur2011crowdforge} uses additional crowd tasks to (iteratively) refine previously submitted tasks; in text acquisition, filtering/ranking~\cite{spirin2012survey,tang2013social,guy2015social,agichtein2008finding,mudambi2010makes,wang2013wisdom,yang2016recommending,siersdorfer2010useful} uses crowd tasks to assess each document's quality and either rank them by quality or filter out low-quality documents.

\begin{figure}[t]\vspace{-1em}
   \centering
    \centering
     \includegraphics[width=.9\columnwidth]{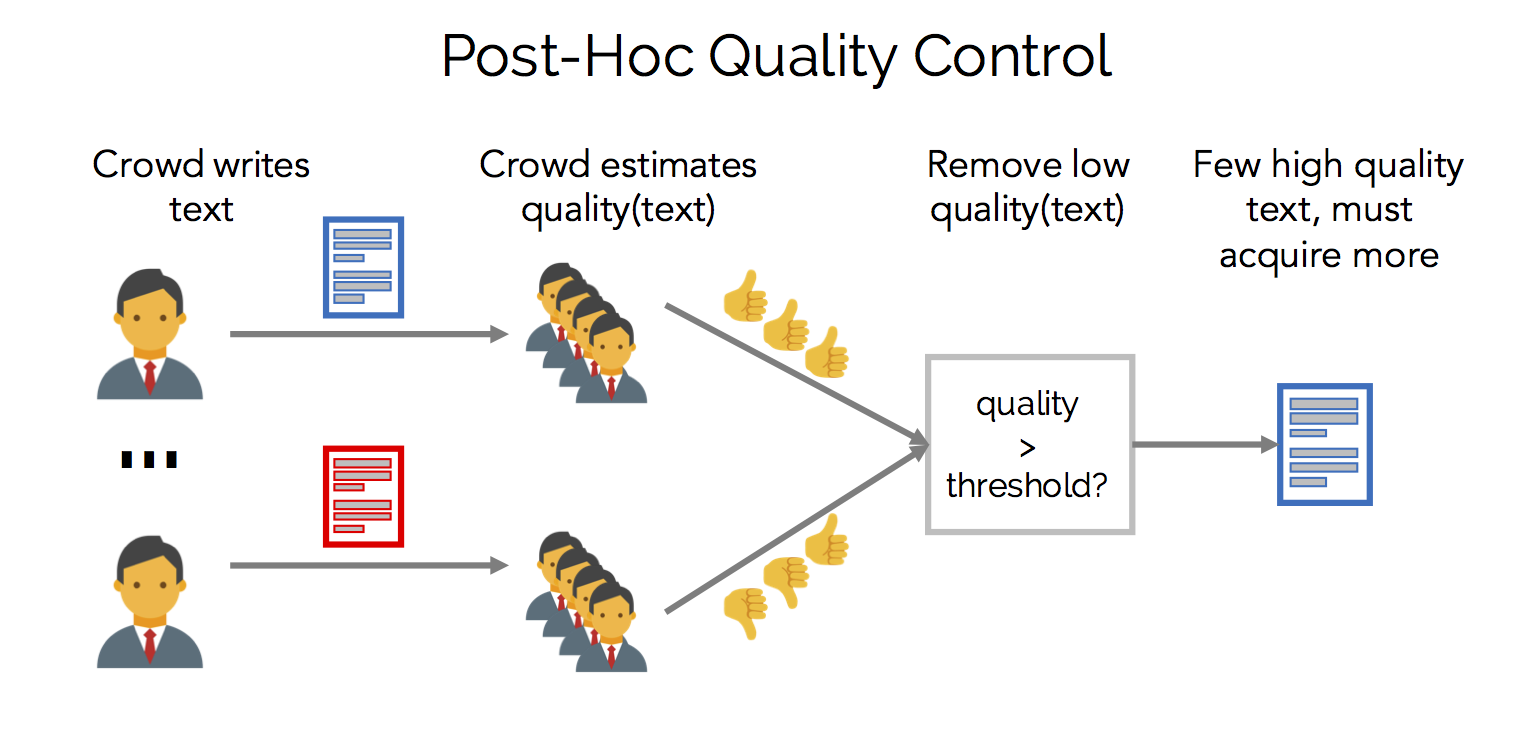}\vspace{-1em}
     \caption{Text acquisition with post-hoc quality control. }
     \label{f:flow-naive}
\end{figure}

\begin{figure}[t]
	\centering
	\includegraphics[width=.9\columnwidth]{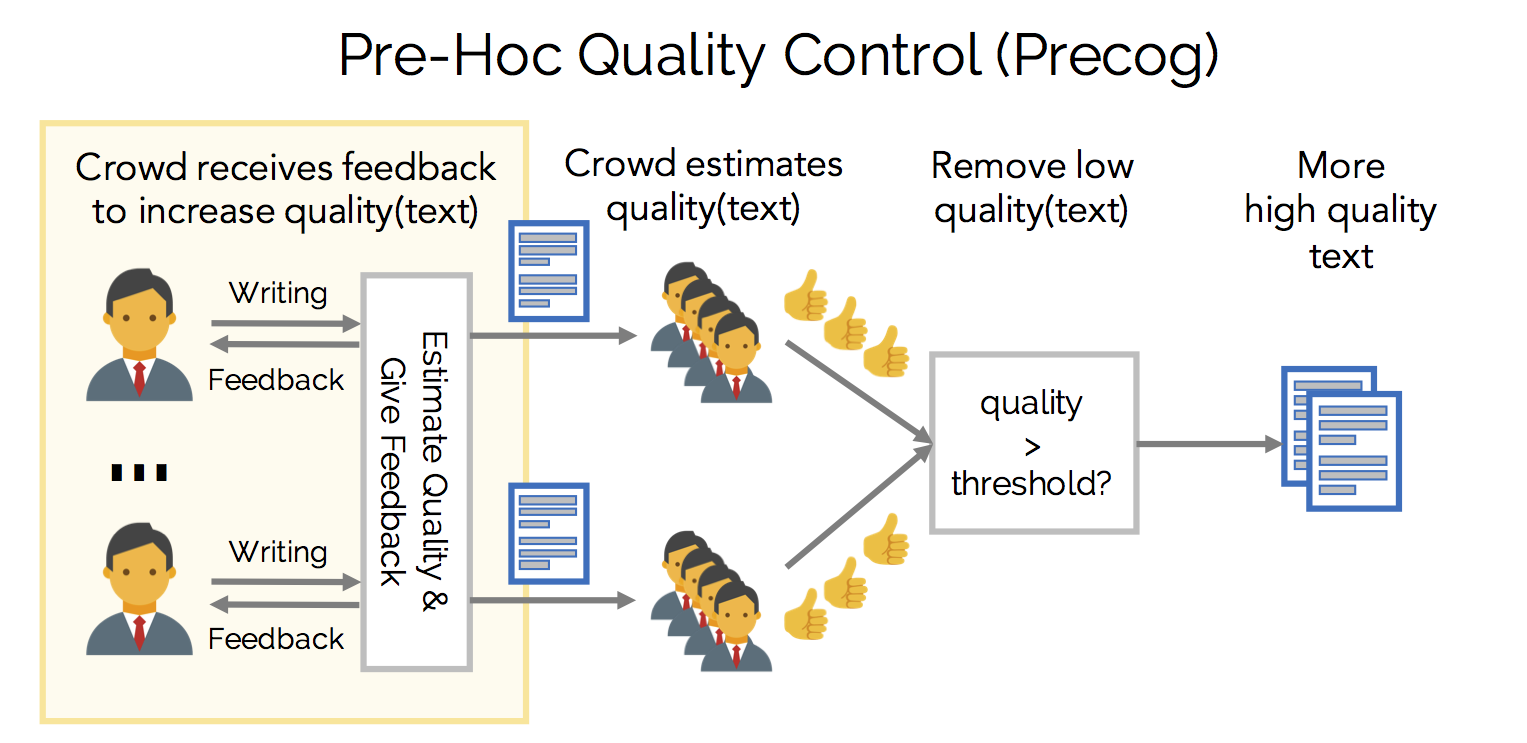}\vspace{-1em}
  \caption{Text acquisition with pre-hoc (beige background) and post-hoc quality control.} \vspace{-1em}
	\label{f:flow-qpushdown}
\end{figure}

In this paper, we argue for \emph{pre-hoc} quality control systems. Pre-hoc quality control occurs before data acquisition and naturally complements many existing \emph{post-hoc} techniques to further improve the final data quality. Figure~\ref{f:flow-naive} illustrates a typical text acquisition workflow: the crowd generates text documents, more tasks are used to estimate the text quality, low-quality documents are removed, and this may ultimately trigger the need to collect more data.  Companies (e.g., Amazon, Zappos) use this post-hoc technique by asking users to assess whether a product review is ``helpful'' or ``not helpful'', and ranks and displays reviews based on this measure.

Figure~\ref{f:flow-qpushdown} augments this workflow with pre-hoc quality control.  The only change is the beige component, which augments the data collection interface (task interface) to estimate the quality of the user's (in this case) text, and automatically provide feedback if the predicted quality is low.    Since good feedback can help the worker improve the text, it naturally improves the quality of the acquired data, and can reduce data acquisition costs.  Furthermore, in some settings where collecting more data is not an option (e.g., less popular products may not have enough users that are willing to, or equipped to, write reviews), it will be more important to apply \prehoc.

In fact, instances of pre-hoc quality control are already commonly used in practice, both in the survey design literature~\cite{groves2011survey} and as form design throughout the Internet. The basic idea is to push \emph{data-quality constraints} down to the data collection interface rather than validate them after data acquisition.
For quantitative attributes, a common data-quality constraint is to ensure values are not out of bounds (e.g., human age should be above~0).  This can be achieved by dynamically identifying these constraint violations and providing feedback to the user.  Similarly, auto-complete may be used to provide feedback about existing categories in order to avoid duplicates when collecting categorical text~\cite{parameswaran2012deco,franklin2011crowddb} (e.g., ice cream flavors, presidents). By tackling low-quality data pre-acquisition, it can reduce or eliminate the need for post-hoc quality control.

Although it's possible to \emph{automatically} perform pre-hoc quality control for simple constraints over simple data types, it is still unclear how this can be achieved for more complex data integrity constraints and data types.  For instance, multi-paragraph \texttt{text} attributes such as product reviews, forum comments, or rental descriptions are particularly challenging for several reasons.  First, the quality measure is continuous (there is no ``perfect document'') and thus hard to identify a ``violation''. Second, it is ill-defined and application-dependent,  thus difficult to specify as a constraint.  Third, it's unclear how to {\it automatically} generate the appropriate feedback text to show the user.  Existing approaches (surveyed in Related Work) focus on syntactic errors such as grammatical mistakes, which cannot help improve the text content, or overly simple models for picking feedback text~\cite{krause2015method}.

To this end, we present \sys\footnote{Similar to precogs in Minority Report~\cite{minorityreport}, who identify and help ``resolve'' low-quality human action in the future, \sys identifies and helps resolve low-quality data before it is submitted in the future.}, a crowdsourced data acquisition system that supports pre-hoc quality control for both simple data types and multi-paragraph text attributes.  It does so by generating feedback or interface changes to help workers improve their data pre-submission.  It can be integrated seamlessly into existing crowdsourcing applications or systems with post-hoc quality control, helping them to further improve quality.  

\begin{figure}[t]\vspace{-2em}
\centering
\includegraphics[width=.95\columnwidth]{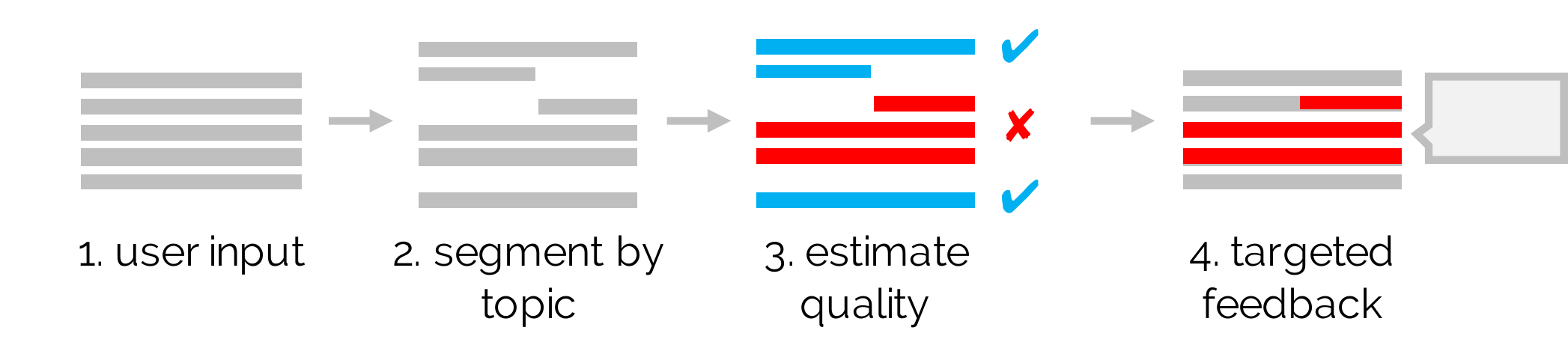}\vspace{-1em}
\caption{The {\it Segment-Predict-Explain} pattern: \sys splits user input into coherent segments; estimates the quality of each segment and the text as a whole; and generates and shows suggested improvements to the user. }\vspace{-.5em}
\label{f:workflow}
\end{figure}

By default, \sys provides optimizations for constraints over numerical and categorical data types, and can be extended with custom optimizations.  Our technical contribution is a pre-hoc feedback system for multi-paragraph text.  As illustrated in Figure~\ref{f:workflow}, we employ a novel {\it Segment-Predict-Explain} pattern to generate customized feedback on an individual segment (rather than document) level. \sys takes long form text from a crowd worker, decomposes it into coherent portions (segments) based on their topics, predicts the quality of each segment, and automatically generates immediate feedback to explain how these segments can be improved. 

The core challenges are to (1) identify a proxy for text quality that is consistent with the downstream application's needs, and (2) to generate {\it effective} feedback text.  We address the former challenge using a {\it data-driven} approach that learns a quality measure from  data that has already been acquired.  For instance, Amazon already has a corpus of high and low-quality reviews, and similarly for other applications.  To build high-quality models, we survey and categorize features from the writing analysis literature into $5$ categories (e.g., readability, informativeness, etc), and implement a representative and extensible library of 47 text quality features.  By default we use this library for learning quality measures from a corpus.

The feedback literature suggests that precise, local feedback is effective~\cite{nelson2007nature}.  Thus, we decompose the text into segments, and for each low-quality segment predicted by the model, we generate segment-specific feedback.  One approach is to simply highlight the low-quality segment and provide generic/static feedback.  Our experiments and prior work~\cite{kulkarni2015} show that this is less effective than a more customized approach.  An alternative is to use existing model explanation algorithms~\cite{ribeiro2016should} to {\it describe} the prediction.  However, it leaves it up to the user to infer specific improvements to make. 

In contrast, we generate {\it prescriptive}, actionable explanations that, if followed, are expected to improve the text.  We define this as the \emph{Prescriptive Explanation} problem, and find that the search space of solutions for the problem is exponential in the number of model features.  Our efficient solution called \sol leverages the structure of random forest models to generate explanations in interactive time.

In addition to evaluating \qp for hard constraints and simple data types, we evaluate \qp's text feedback through extensive MTurk experiments on two real application domains---product reviews and rental host profiles.  \sys is easily extended to new domains, and increases the number of high-quality documents by $\geq 2\times$ compared to not using pre-hoc techniques. We further show that \sys's unique approach to combining {\it prescriptive} explanations and segment-level feedback improves text quality by $14.3\%$, and over $3\times$ better than a state-of-the-art feedback system~\cite{krause2015method}.  To summarize our contributions:
\begin{itemize}[leftmargin=*, topsep=0mm, itemsep=0mm]
  \item We present the argument for \prehoc and present its unique advantages as well as the challenges for multi-paragraph text.
  \item The design and implementation of \sys, which supports \prehoc for constraints over simple data types and quality measures over text and open-ended attributes. 
  \item A data-driven approach to estimate quality for text attributes, including a categorization and implementation of $47$ text quality features from a survey of the literature.
  \item We define the {\it Prescriptive Explanation} Problem to provide actionable feedback for text acquisition.  The problem is exponential and we present an efficient solution that leverages the structure of random forest models to generate high-quality feedback in interactive time.
  \item Extensive MTurk experiments on two real-world domains with different quality measures: helpfulness for Amazon product reviews and trustworthiness for AirBnB housing profiles.  \sys, which is complementary to \posthoc techniques, collects $\geq 2\times$ high-quality documents for the same budget as no feedback, and improves text quality by $14.3\%$ on average.

\end{itemize}

\section{{\large \sys}: a Precog System}
\label{s:arch}

As described in the introduction, \sys seeks to optimize the data collection interface in order to improve the quality of the collected data and ensure data quality constraints.
In this section, we first describe how users express \qp quality control for common data integrity constraints, as well as quality scores on a crowd-sourced table.
Quality scores are intended for attribute values for which the definition of quality defined as a continuous measure to be improved, rather than a boolean constraint, and provides the framework for which we implement a model-based feedback system for performing \qp on text attributes (Section~\ref{s:sep}).

\subsection{Pushing Data Constraints to the Interface}

\sys extends existing crowdsourced databases that contain crowdsourced and non-crowdsourced base relations;
a crowdsourced table~\cite{franklin2011crowddb} represents a subset of all possible records that may be stored in the table and the task is to acquire records to insert into the table.
\sys uses existing techniques to generate forms for crowd workers to fill out, and the form contents are inserted as new records into the corresponding crowd table.
For instance, Amazon product reviews and users may be modeled using the following crowd-based DDL statements.
The first states that user information is collected from the crowd (of Amazon users) and that the username must be unique.
The second states that a review is written for a given product in the \texttt{products} table, and contains a numerical rating as well as the text of the review.
For the sake of exposition, \texttt{product\_id} is the textual name of the product.
The final \texttt{FEATURE} table \texttt{review\_feats} is used in the later sections to represent the features extracted from the value of the primary key (\texttt{review}).  For instance, \texttt{len FEATURE len\_extracton} defines the feature returned by the user-defined function \texttt{len\_extracton}.
{\small
\begin{verbatim}
  CREATE CROWD TABLE users (
    id autoincrement primary key,
    username text UNIQUE,
    age int CHECK age > 0 AND age < 100,
    CHECK(username matches \w+)
  );
  CREATE CROWD TABLE reviews(
    id autoincrement primary key,
    product_id text,
    rating int CHECK rating > 0 AND rating <= 5,
    review text,
    QUALITY SCORE qualreview qual_udf(review),
    FOREIGN KEY product_id REF products(id)
  );
  CREATE FEATURE TABLE review_feats(
    review text primary key references reviews.review,
    topics FEATURE topic_extractor,
    len FEATURE len_extracton,
    ...
  );
\end{verbatim}
}

In addition to boolean constraints such as domain, foreign key, and uniqueness constraints, \sys also supports {\it quality scores}.  In contrast to typical integrity constraints, which will reject an inserted record that violates the constraint,  \sys seeks to maximize its value.  For instance,  \texttt{qualreview} seeks to maximize the quality score as defined by \texttt{qual\_udf}.    This provides the functionality for our automatic \prehoc system for free-form text attributes.

The rest of this subsection describes the \texttt{DDL} statements that users can use to specify feedback and interfaces for \qp quality control. These statements complement existing task interface specifications that prior crowdsourcing systems~\cite{marcus2011human,franklin2011crowddb,parameswaran2012deco} use for task generation by providing a way to augment them for data integrity constraints.

\begin{figure}[tb]\vspace{-2em}
\centering
\hspace*{-0.025\columnwidth}
\includegraphics[width=1.05\columnwidth]{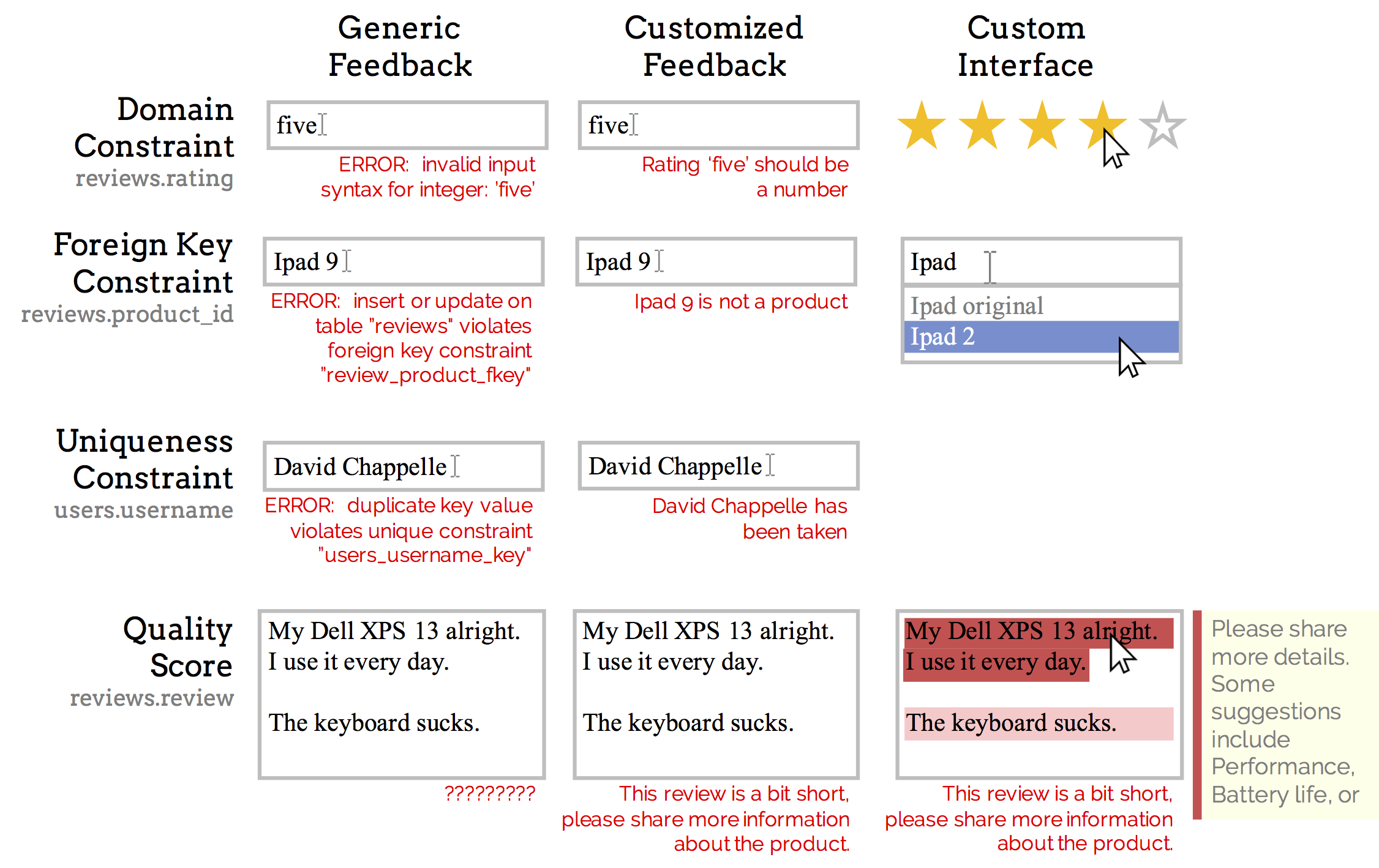}
\caption{Examples of three levels of \qp quality control for four classes of data integrity constraints. 
}
\label{f:feedbackinterfaces}
\end{figure}

\stitle{Overview: }
In contrast to naive form validation, which simply rejects user inputs with an error message, \sys seeks to accommodate iterative improvements through feedback interfaces.
Figure~\ref{f:feedbackinterfaces} summarizes \qp into three levels based on the amount of customization needed by the developer.  The default simply renders feedback generated from database constraint violations on tuple insertion (left column).  Developers commonly implement {\it explanation functions} to generate more user-friendly feedback (middle column).  Finally, the most sophisticated may change the input element itself in order to constrain or fully customize the feedback (right column).  

Below, we describe how developers can express the three levels of \qp quality control for domain, foreign-key, uniqueness, and quality score constraints in Figure~\ref{f:feedbackinterfaces}.  

\stitle{Generic Feedback: }
\sys automatically generates feedback based on the error message that the underlying database generates when the \texttt{INSERT} violates a constraint.  The left column shows the feedback interface generated by default.  Although they are interpretable for simple constraints such as domain violations, the language for the uniqueness violation requires database familiarity and may not be accessible to non-technical experts.  Since the quality score is not a boolean constraint, feedback is simply not generated for it\footnote{\scriptsize Note that the developer may express a \texttt{CHECK} constraint and the database can generate an (indecipherable) error message.}.  As constraints become more complex, there is a need for customized messages.

\stitle{Customized Feedback: } 
\sys provides a \texttt{DDL} for developers to customize feedback.  A developer first defines an explanation function that takes as input the list of attribute names and values for which the constraint is defined for (in order to support multi-attribute constraints) and the error message, and returns a string that is shown as feedback.  In these examples, we simply define a python function.  The developer then binds an explanation fuction to the appropriate constraint.
{\small
\begin{verbatim}
  def exp_func(att1, val1, ..., attn, valn, err=None):
    return "custom error message"
  
  CREATE EXPLANATION <func> ON <table>(<att1>,..<attn>) 
  FOR   <CONSTRAINT NAME> USING <explanation function>;
\end{verbatim}
}
Below is the specification to customize the feedback for a numeric domain constraint\footnote{\scriptsize Note that databases automatically generate names for almost all integrity constraints.  Some constraints, such as domain constraints, are registered as syntax errors.  For these, \sys generates default names of the form \texttt{<table>\_<attribute>\_<type>}.}.  Note that the same explanation function is used for domain constraints on \texttt{reviews.rating} and \texttt{users.age}.  
{\small
\begin{verbatim}
  def numeric_exp(att, val, err): 
    return "%s: `%s' should be a number" % (att, val)
  
  CREATE EXPLANATION ON reviews(rating) 
  FOR reviews_rating_domain USING numeric_exp;
  
  CREATE EXPLANATION ON users(age) 
  FOR users_age_domain USING numeric_exp;
\end{verbatim}
}
Similar functions can easily be written for the foreign-key and uniqueness constraints in Figure~\ref{f:feedbackinterfaces}:
{\small
\begin{verbatim}
  def product_exp(att, val, err): 
    return "%s is not a product" % val
  
  def unique_exp(att, val, err): 
    return "%s has been taken" % val
\end{verbatim}
}

For text attributes, the explanation function is slightly different, which is defined on a \texttt{FEATURE} table. An example will be shown in Section~\ref{s:fef}.

\vspace{.5em} 

Although these user defined functions are powerful enough to support arbitrary analysis of an attribute value, such an approach is difficult to compose and extend, and the feedback is still limited to the entire attribute value.  In many cases, such as text attributes, it is desirable to provide feedback for specific segments of the text value.   For this, we next introduce \texttt{DDL} statements to specify custom interfaces.

\stitle{Custom Interface: } 
Fully customizing the interface component is useful in order to directly prevent users from submitting invalid attribute values.  For instance, we might replace the \texttt{rating} domain constraint with five stars similar to Yelp and other social websites.  However, we may use a slider if for larger cardinalities. We assume that the interface is a javascript function (say, as an AngularJS~\cite{darwin2013angularjs} or ReactJS~\cite{fedosejev2015react} component); the constructor takes as input a \sys-provided \texttt{getFeedback} method that retrieves feedback from the \sys server.  Developers can bind the interface to an attribute using a \texttt{CREATE INTERFACE} statement.  For example, the following specify the star interface for \texttt{rating} and the autocomplete interface for \texttt{product}:
{\small
\begin{verbatim}
  CREATE INTERFACE ON reviews(rating)
  USING "stars" FROM "interfaces.js" 
  AND explanation_function;

  CREATE INTERFACE ON reviews(product_id)
  USING "autocomplete" FROM "interfaces.js"
  AND explanation_function;
\end{verbatim}
}
It addition, custom interfaces can be used to provide feedback that goes beyond textual feedback (e.g., visualizing distributions of common numerical values), or that is at a finer granularity than for the entire attribute.  For instance, the bottom row of Figure~\ref{f:feedbackinterfaces} illustrates fine-grained feedback in the form of both highlighted text and text feedback for individual segments that the user has written for \texttt{reviews.review}. Section~\ref{s:sep} describes the \isep pattern that helps developers easily customize interface for text attributes.

\section{Segment-Predict-Explain}\label{s:sep}

The challenge with directly developing \qp quality control for text is that the quality score and explanation function is difficult to express as a concrete function, and they must be customized for the application domain. To address this issue, we present a \isep pattern that reduces the developer's efforts by allowing them to express the quality score in terms of model features by defining a \texttt{FEATURE} table, and to define explanation functions over features of the text attribute.   
Our design is informed by the writing analysis and feedback literature, which emphasizes the value of providing immediate feedback~\cite{kulik1988timing}, as well as fine-grained feedback for specific portions of the text~\cite{choudhury2016,singh2012automated,rivers2014automating}, as is common in coding environments.

Existing feedback approaches are not directly applicable for \sys.  Crowd-based feedback is effective, but can take 20 minutes to generate feedback~\cite{kulkarni2015} and are essentially post-hoc because they create new crowd tasks to refine previously submitted ones.  Automated approaches such as auto-graders primarily focus on predicting quality rather than generating feedback~\cite{valenti03,farra2015scoring,attali2004automated,madnani2014explicit}; others are limited to syntactic analysis~\cite{word,foxtype,googledoc}, or generate overly simple writing feedback~\cite{krause2016interacting,boomerang,biran2014justification}.
In the rest of this paper, we use the term {\it document} to refer to the value of the acquired text attribute.

\stitle{\sep:} Based on these observations, \sys automatically identifies low-quality portions of a document, and generates feedback to help improve the identified issues.
In order to generate targeted feedback, \sys automatically identifies topically coherent portions and {\it segments} the document in order to analyze each segment individually.  For this, we use TopicTiling~\cite{riedl2012topictiling}, a sliding window-based segmentation algorithm that computes the dominant topics within the window using LDA~\cite{blei2003latent}.  When the topic within the window changes significantly, then TopicTiling creates a new segment.  \sys is agnostic to the specific segmentation algorithm, and developers can use their own.

Rather than define a concrete quality measure, \sys automatically learns the quality measure from a training corpus that contains documents along with their quality labels (for the entire document, not each segment).
We learn this quality measure by training a random forest model that {\it predicts} the quality of individual text segments.
We believe our assumption about the availability of a training corpus is reasonable in data acquisition settings, because such quality labels are already gathered in order to rank documents (e.g., Amazon helpful/unhelpful reviews, Reddit comment up/down votes).
We describe this in Section~\ref{s:predict}.

Finally, \sys\  {\it explains} why segments were predicted as low quality by  selecting the feedback that is most relevant to changing the segment into a high quality prediction.  To do so, we develop a novel perturbation-based analysis to identify the combination of features that, when changed, will most likely reclassify the text as high quality.  We then map these feature combinations to explanation functions that are executed to generate the final set of feedback text (Section~\ref{s:explain}).

\begin{figure}[t] \vspace{1em}
\centering
\includegraphics[width=1\columnwidth]{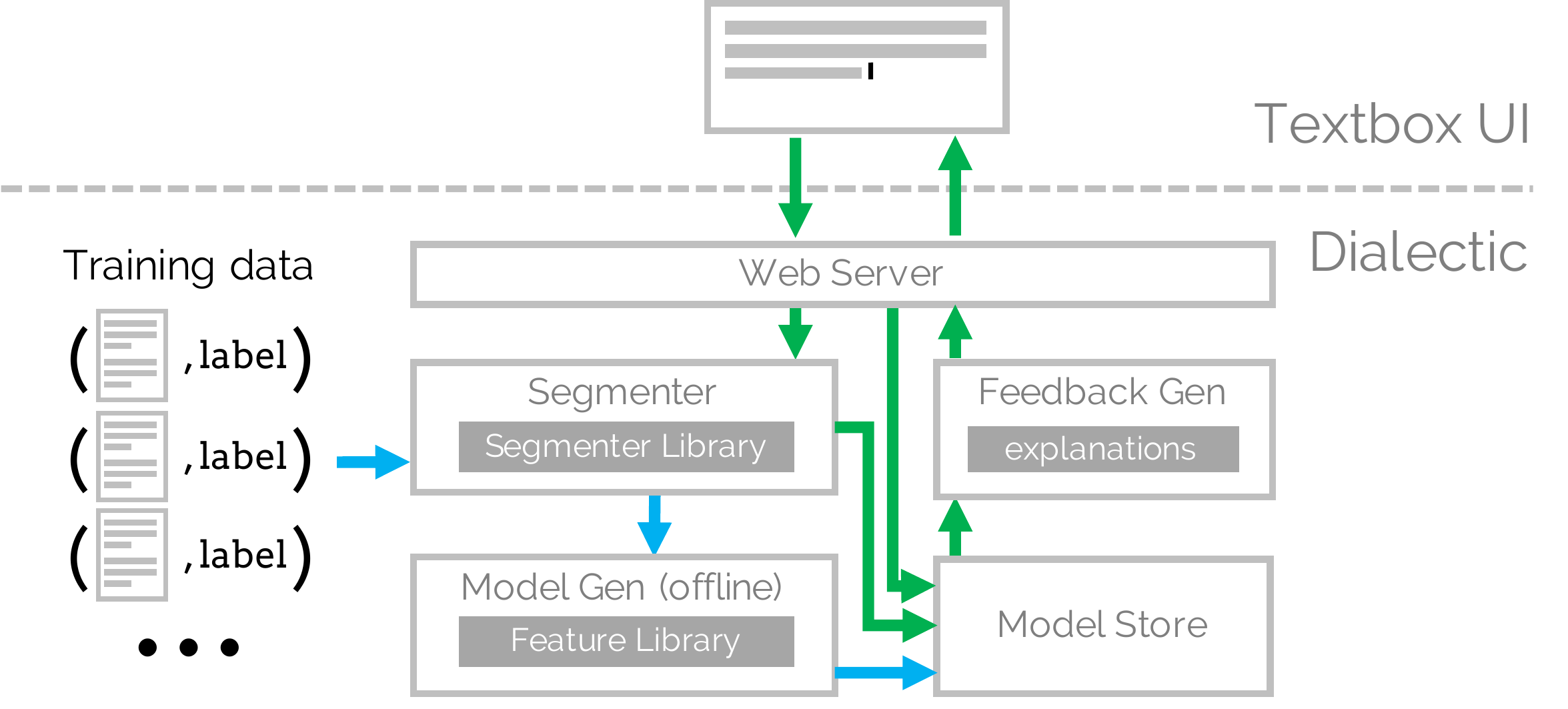}
\caption{\sys architecture. {\bf\purple{Purple}} arrows show the feedback process for hard constraints.  
  The \isep component has a beige background: {\bf\blue{Blue}} arrows depict the offline training and storage process and 
  {\bf\green{Green}} arrows depict the online execution flow when a user submits. }
\label{f:arch}
\end{figure}

\stitle{User-facing Interface:}  The custom interface column for the quality score in Figure~\ref{f:feedbackinterfaces} depicts the \qp interface in action.  
The user writes a product review in the textbox; the content is sent to the \sys backend via \texttt{getFeedback()}.  The backend splits the review into coherent segments, identifies the low-quality segments, and generates document-level feedback.  The document-level feedback is shown to the user, and the low-quality segments are highlighted as light red in the interface.  Finally, when the user hovers over a highlighted segment, more targeted feedback helps explain why it was identified as low quality and how it could be improved.

\stitle{Architecture:} Figure~\ref{f:arch} depicts the system architecture. 
For hard constraints ({\bf\purple{Purple}}), user inputs are sent to the database, which checks that the input satisfies the integrity constraints.
On violations, the feedback generator creates custom feedback (if specified in a \texttt{DDL} statement) and the default or customized interface displays the feedback.

The \isep component consists of offline and online components.   The offline components ({\bf\blue{blue}} arrows) take as input a corpus of training data in the form of user generated text documents and their labels---for instance, Amazon product reviews may be labeled by the ratio of ``helpful'' and ``unhelpful'' votes. The  {\it Segmenter} first splits each document into segments. The {\it Model Generator} then trains two classification models to predict the quality of a user's overall text submission as well as its constituent segments; these are cached in the {\it Model Store}. 

The online components ({\bf\green{green}} arrows) send the contents of a text input widget, along with an optional corpus name, to the webserver.  \sys uses the models in the {\it Model Store} to identify whether the entire document and/or segments generated by the {\it Segmenter} are low quality.  The {\it Feedback Generator} then constructs feedback explanations for the low quality text, which are returned and displayed in the widget.

\section{Predict}
\label{s:predict}

\sys takes as input a training corpus of documents and document-level quality labels, and trains two models---document-level and segment-level prediction models---in order to provide document-level and fine-grained segment-level feedback. Both are important because they address different text quality factors. The document level feedback provides a global quality assessment.  For instance, consider a document that contains a single segment---the segment may be high quality but the overall document is too short and is missing text for other topics.  In contrast, segment level feedback is needed in order to provide specific, actionable suggestions that may not be evident at the document level.

Clearly, document quality assessment is a well-studied area. In this section, we describe our approach towards in-depth semantic feedback. We first describe our extensible feature library that consolidates text features across literature in social media text analysis, essay grading, language psychology, and data mining research communities. As compared to other features libraries such as LIWC, \sys's main advantage is a high-concentration of data-driven features (topic modeling, jargon usage, text similarity measures) that are trained to fit each developer's unique corpus. Further, developers can easily extend the library with custom features.

Based on this library, we develop document-level and segment-level prediction models.  The key challenge is that training data only contains quality labels for entire documents (e.g., helpfulness for the full review), and it is unclear how to leverage them for training a segment-level model.  We describe our experiments that show that it is possible to use these labels as a proxy for individual segments.

\subsection{Feature Library for Text Quality}

\begin{table*}[ht] 
\hspace*{-.1in}
\centering
\small
  \begin{tabular}{l|l|m{42em}}
    {\textbf{Category}} &  {\textbf{\#}} & {\textbf{Description}} \\ \hline
    Informativeness  & 8 &  mined jargon word and named entity stats~\cite{Hu2004}, length measures (word, sentence, etc. count)\\ \hline
    Topic & 5 &  LDA topic distribution and top topics ~\cite{blei2003latent}, entropy across topic distribution \\ \hline
    Subjectivity  & 15 &  opinion sentence distribution stats ~\cite{Hu2004}, valence, polarity, and subjectivity scores and distribution across sentences ~\cite{ghose2011estimating,gilbert2014vader,loria_2014}, \% upper case characters, first person usage, adjectives\\ \hline
     Readability and Grammar & 15  &  spelling errors ~\cite{kelly_2017}, ARI, Gunning index, Coleman-Liau index, Flesch Reading tests, SMOG, punctuation, parts of speech distribution, lexical diversity measures, LIWC grammar features\\ \hline
    Similarity & 4 &  various TF-IDF and top parts of speech comparisons with sample of low and high utility documents \\ \hline

    \end{tabular} 

  \caption{Summary of feature library for text quality.} \vspace{-.5em}
  \label{t:features}
\end{table*}



Existing automated writing feedback tools primarily focus on syntactic, simple errors ~\cite{word,foxtype,googledoc}. However, recent study shows the promise of translating semantic features to textual feedback~\cite{krause2015method}. Our goal is to provide the foundation for such content-specific semantic feedback by surveying and categorizing features from the writing analysis literature. 

To this end, we performed a survey of literature spanning of social media text analysis~\cite{ma2017self,tan2016winning,siersdorfer2010useful,ghose2011estimating,liu2007low}, essay grading~\cite{valenti03,farra2015scoring,attali2004automated,madnani2014explicit}, deception detection~\cite{markowitz2016linguistic,drouin2017linguistic}, and information retrieval~\cite{Hu2004,spirin2012survey}. Our contribution is to curate the subset of these features that can be generalized across text domains to improve writing quality, categorize them (Table 1), and integrate them into an open source feature library\footnote{Available at \url{http://cudbg.github.io/Dialectic}}. 
This groundwork reduces the task of applying \sys to new domains. 
The primary features that we do not include are those that rely on application metadata such as the worker's history or location, which may be predictive of quality but not related to the writing content, and cannot be mapped to actionable writing feedback. 

We identify five main categories across the existing literature (Table 1). The first category, \emph{Informativeness}, highlights trends across existing literature that show that both general length measures ~\cite{ma2017self,attali2004automated,liu2007low,siersdorfer2010useful,krause2015method} as well as domain-specific jargon are highly predictive of quality ~\cite{markowitz2016linguistic,liu2007low,ma2017self}. We implement a variety of length measures, and use the Apriori algorithm~\cite{Agrawal1994} to mine jargon based on the training data inputted into \sys, and identify its distribution across the sentences of an input document. Moreover, there have been many successful attempts to use \emph{topic} distributions to predict quality~\cite{ma2017self,liu2007low,ma2017self}. While such approaches are often supervised in nature, requiring a manual topic ontology~\cite{ma2017self,mcauliffe2008supervised}, we use LDA~\cite{blei2003latent} because it is unsupervised and can be quickly trained on any corpus without any cost to the developer. Furthermore, while most approaches simply use the distribution of topics as a feature~\cite{ma2017self,mcauliffe2008supervised}, \sys computes several summary statistics (entropy, topic ID and probability of top-K topics, ranked by probability) not used in prior work that prove highly predictive in our experiments. \emph{Subjectivity} assesses user bias using a variety of features ranging from sentiment analysis~\cite{ghose2011estimating,gilbert2014vader,loria_2014} to pronoun usage~\cite{pennebaker2015development}. \emph{Readability/Grammar} is an aggregate of syntactic features shown predictive across multiple domains~\cite{siersdorfer2010useful,krause2015method,markowitz2016linguistic,drouin2017linguistic}. Finally, the \emph{Similarity} category reflects how many quality prediction approaches compare the input document to a gold-standard of text~\cite{kim2006,krause2015method}. We compute a variety of similarity measures between the input document and a sample of high and low quality documents--using both the simple TF-IDF measure used in prior work~\cite{kim2006} as well as occurrences of popular parts of speech appearing in a document (i.e top-K nouns in unhelpful documents that appear).

\subsection{Document-level Prediction}
We now describe the prediction model we use for document-level prediction. Once a library of features are given, the document-level prediction turns to be a typical classification problem. We choose a random forest classifier, which has been shown effective in existing work~\cite{ghose2011estimating}, and select features using the recursive feature elimination algorithm~\cite{guyon2002gene}.  


Our model performs competitively with prior work~\cite{ghose2011estimating}. The prior work predicts the quality of Amazon DVD, AV player and Camera reviews with  $83\%$ accuracy;  \sys's default model on the same setup predicts at $85\%$ accuracy---the slight improvement is due to the additional features in the topic and similarity categories from other literature (Table~\ref{t:features}). \sys also achieves $79\%$ accuracy at predicting if an Airbnb profile is above or below median trustworthiness, using trustworthiness data from~\cite{ma2017icwsm}. We validated generalizability of the model to domains not covered in prior work by evaluating it on reddit comments from the AskScience subreddit\footnote{{\small https://www.reddit.com/r/askscience/}} and predicted comment helpfulness on an evenly balanced sample with $80\%$ accuracy\footnote{{\small We define $>1$ net up-votes as helpful and $\le1$ as unhelpful.}}. 



\subsection{Segment-level Prediction}
There are two challenges in training a segment-level prediction model.  The first one is how to split a document into segments.  Although there are numerous segmentation algorithms, we describe the rationale for the choice of using a topic-based segmentation algorithm.  The second challenge is to determine how the available document-level labels can be used for training segment-level quality.

\stitle{Segmentation:} Contributor rubrics across many social media services are structured around topics~\cite{yelpguidelines,amazonguidelines,wikiguidelines}, and psychology research suggests that mentally processing the topical hierarchy of text is fundamental to the reading process~\cite{hyona2002individual}.  Thus, \sys segments documents at topic-level units.  To this end, we use a technique called TopicTiling~\cite{riedl2012topictiling}, an extension to TextTiling~\cite{hearst1997texttiling}. It uses a sliding window to compute the LDA~\cite{blei2003latent} topic distribution within each window and create a new segment when the distribution changes beyond a threshold.  TopicTiling outperformed other topic segmenters~\cite{misra2011text,Wu2011} in terms of their WindowDiff score~\cite{pevzner2002critique} as compared to a hand-segmented test corpus of 40 documents.

Moreover, \sys also makes it easy for developers to add custom segmentation algorithms.  Given a small test corpus of pre-segmented documents, \sys can benchmark the algorithms and recommend the one with the highest WindowDiff score.

\vspace{.5em}

\stitle{Document Labels for Segments:} Despite generating topically coherent segments, we lack quality labels for training the predictive model at the segment level. One solution is to manually label the generated segments, but this will be very costly and time-consuming. We observe that document quality is sufficiently correlated with segment quality, and a document's label can be used to label its segments as training data for a segment classifier. The key insight is that the predictive model is robust to noisy labels. Although there might be a number of segments mislabeled, the model can tolerate their impact well and  achieve good performance.

We tested this hypothesis by running an experiment, using an existing corpus of Amazon reviews~\cite{amazondataset}. We compared a segment binary classifier trained under this assumption with human evaluation. Specifically, we ran a crowdsourced study to label $500$ Amazon segments ($250$ drawn from helpful reviews, and $250$ from unhelpful reviews), with human helpfulness labels (the median segment length of a review is 3). We trained workers on a separate sample of segments, along with explanations of why each segment was helpful or unhelpful. We then randomly assigned each worker $50$ segments to label, and collected labels until each segment had $\geq 3$ labels, and determined the final label of each segment using the Get Another Label algorithm~\cite{sheng2008get}. 

We then computed pairwise accuracies between the document labels, classifier predictions, and crowd labels: $71.1\%$ (Classifier predicting Crowd Label), $72.5\%$ (Classifier predicting Document label), and $69.5\%$ (Document label predicting Crowd Label). The consistent results between all three comparisons suggest the efficacy of the segment-level classifier, and our end-to-end experimental results suggest that the predictive model is effective at providing segment level feedback.
Nevertheless, more studies are needed to fully evaluate this hypothesis across other text domains and document lengths. We defer this to future work.

\section{Explain}
\label{s:explain}

We describe how \sys automatically generates feedback text for low-quality text.  This problem is challenging because we must analyze potentially arbitrary text content.  Our approach is inspired by existing feedback systems---model features act as signals to identify text characteristics that the worker should change.  
We first introduce the {\it Prescriptive Explanation} problem, which assigns responsibility to each model feature proportional to the amount that it will contribute to improving the predicted text quality.  We then use explanation functions to transform the most responsible features into prescriptive feedback for the user.

\subsection{Problem Background}\label{s:explain-related}

Our problem is closely related to model explanation, which  generates explanations for a model's (mis-)prediction.  The classic approach is to use simple, interpretable models~\cite{caruana2015intelligible,letham2015interpretable,ustun2016supersparse} or to learn an interpretable model using the training data near the test point~\cite{ribeiro2016should}.  However, it still leaves it up to the user to infer specific improvements to make.

Feedback systems are typically based on outlier detection~\cite{krause2015method}.  They pre-compute the ``typical'' values of each feature in the high quality corpus, then identify the ``atypical'' outliers in the test data's feature vector (e.g., a feature whose value is 1.5 standard deviations from the mean).  Features are individually mapped to pre-written feedback text~\cite{krause2016interacting,boomerang,biran2014justification}.  Unfortunately, this procedure is not effective for non-continuous or low cardinality features such as one-hot encoded features (e.g., each word is represented as a separate binary feature) common in text analysis.

Further, their analyses are per-feature and don't account for multi-feature interactions.  
Consider a review consisting of a long, angry diatribe about customer service. In isolation, existing approaches may find that the length is large and suggest reducing it, and that the emotion is high and suggest reducing it.  However, such systems would not recognize that the review can be most improved by simultaneously reducing the emotion in the text and including more product details that ultimately {\it increase} the length.

Ultimately, existing feedback and explanation approaches are descriptive of the prediction, rather than prescriptive of the changes that must be made. Although the data cleaning literature has proposed ways to prescribe data cleaning operations~\cite{chalamalla2014descriptive}, they are not applicable for text attributes.  We directly address this problem by selecting multi-feature explanation functions to prescribe improvements to the user's text.

\subsection{Feature Explanation Functions}\label{s:fef}

Section~\ref{s:arch} introduced explanation functions that can take as input features in a \texttt{FEATURE} table whose primary key references the desired text attribute.
We now formally define these \emph{feature-oriented explanation functions}  (FEFs) and provide examples used in the experiments.

Let $\mathcal{F}$ be the set of $n$ model features, and $f_i$ denote the $i^{th}$ feature.
An FEF $e: \mathbb{R}^{|\mathbb{F}|}\rightarrow text$ maps the feature vector for a subset of features $\mathbb{F} \subseteq \mathcal{F}$ to feedback text.  Intuitively, the FEF should be executed if its list of features $\mathbb{F}$ can take ``highly responsibility'' for improving the quality score.  \sys can automatically control the generated feedback by reallocating responsibility.

In practice, an FEF takes as input a list of features, as well as the text document and the full feature vector, and returns feedback text.  Recall the feedback in the custom \sys interface in Figure~\ref{f:feedbackinterfaces}, it identifies that the segment is short on details and suggests new topics.   The following snippet sketches the {\it Not Enough Detail} function in our evaluation.  If the features \texttt{topics}, \texttt{featureCnt}, and \texttt{textLen} have high responsibility, then it will be called to recommend new product features that the worker should mention in the review; the recommendations are dynamically selected based on the text's topic distribution (\texttt{topics}) and the number of product features detected (\texttt{featureCnt < 10}):
{\small
 \begin{verbatim}
def notEnoughDetail(topics, featureCnt, textLen, 
                    text="", feats=[]):
  if featureCnt < 10 and textLen < threshold:
    return "Try adding information about: " +
    suggest_new_prod_feats(topics, text, feats)
  ...
\end{verbatim}}
We note that existing feedback systems~\cite{krause2016interacting,krause2015method,biran2014justification} implicitly follow this model, however they bind individual features to static strings.   In contrast, \sys supports feature combinations and can dynamically generate feedback based on the input text.  Although developers can easily implement their own FEFs, \sys is pre-populated with $5$ FEFs that work across the two application domains used for evaluation.

\subsection{Problem Statement}

\begin{figure}[t]
   \centering
   \includegraphics[width=.6\columnwidth]{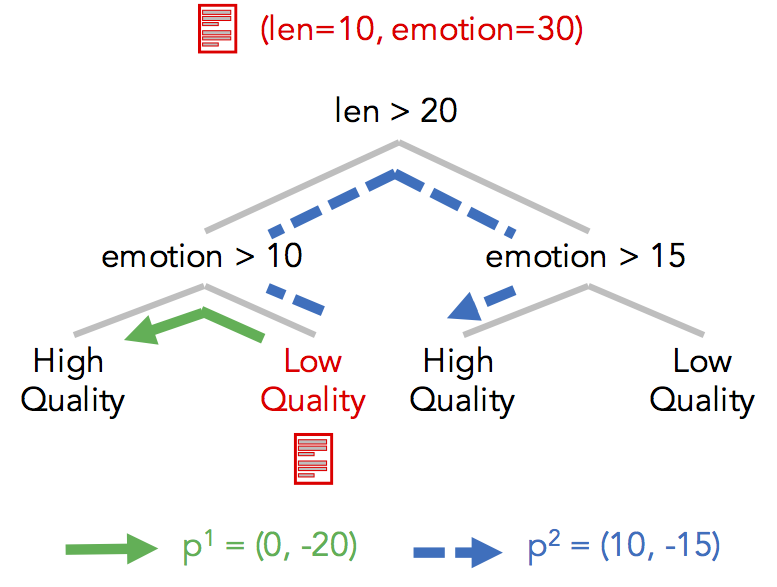}
   \caption{ Assigning responsibility to perturbations.  The paths go from the document's current low quality classification to a high quality classification.
   The green path ($p^1$) must at least reduce emotion by $-20$; the blue path ($p^2$) must at least increase length by $10$ and at least reduce emotion by $-15$.}
   \label{f:ex-perturb}
\end{figure}

\stitle{Intuition: } Figure~\ref{f:ex-perturb} depicts the main intuition behind the problem and our approach.  Consider a single tree in a random forest, consisting of decisions on two features, \texttt{len} and \texttt{emotion}.  \sys uses the feature library to transform the input text into a feature vector of \texttt{(len=10, emotion=30)}, and is thus classified as low quality.  

When we consider a user's edits, they are desirable if the edits will improve the document's quality---in other words, if it will cause the document to be reclassified as high quality.  In this example, there are two ways to {\it perturb} the feature vector: by reducing the \texttt{emotion} feature by at least $20$, or by increasing the \texttt{length} by at least $10$ and reducing the \texttt{emotion} by at least $15$.

Thus, it is clear that the \texttt{emotion} should be assigned a greater responsibility because there are more branches for which changing its value will contribute to a better classification.   In general, we must account for the amount that a feature must be {\it perturbed}, and the number of other features that must also be {\it perturbed}, in order to improve the classification.  A similar approach is applicable for regression models as well, where increasing the continuous prediction assigns the perturbation more responsibility.

\stitle{Setup:} 
  Let $d \in \mathbb{R}^{n}$ be a data point (text document or segment) represented as a feature vector, where $d_i$ corresponds to the value of $f_i$.  For instance, $\mathcal{F}$ may be the text features described above, and a data point corresponds to the extracted text feature vector.  A model $M:\mathbb{R}^{n}\rightarrow\mathbb{N}$ classifies a data point as $M(d) \in \mathbb{N}$, and a utility function $U: \mathbb{N}\rightarrow\mathbb{R}$ maps a label to a utility value.  For instance, in a binary classification problem $U$ may return $1$ if the input is ``high quality'' and $0$ otherwise; in a regression model, $U$ may be the identity function.  

  A perturbation $p \in \mathbb{R}^{n}$ is a vector that modifies a data point. $p_i\ne0$ if $f_i$ is perturbed, otherwise $p_i=0$.  We assume that the domains of the features have been normalized between $[0, 1]$.

\stitle{Responsibility: } 
Our goal is to identify feature subsets of the test data point $d$ that, if perturbed, will most improve $d$'s utility\footnote{No feedback needed if data point already has high utility.}.  
To do so, we first define the {\it impact} $I(d,p)$ for an individual perturbation $p$ as the amount that it improves the utility function discounted by the amount of the perturbation $\Delta(p)$ and the model's prediction confidence $C(d+p) \in [0, 1]$.  
{\small
\begin{align*}
  I(d,p) &= \frac{U(M(d+p)) - U(M(d))}{\Delta(p)}\times C(d+p)
\end{align*}
} %
$C$ can be chosen based on the model---for a random forest, we define $C$ as the percentage of trees that vote for the majority label.
The discount function $\Delta$ can be similarly defined in multiple ways.  

For instance, consider $\Delta(p) = |p|_2$,  the L2 norm of the perturbation vector. It will cause the impact function to converge to $0$ as the perturbations become larger.  Consider the perturbations $p^1, p^2$ in Figure~\ref{f:ex-perturb}. 
Assuming that $C()=1$, $p^1$'s impact on the input document is $I(d, p^1) = \frac{1-0}{20}\times 1 = 0.05$, whereas $p^2$'s impact is $I(d, p^2) = \frac{1-0}{10^2+15^2}\times 1 = 0.055$.

However, there can be an infinite number of perturbations that all improve the utility---which should be selected?  In this work, we restrict the analysis to perturbations that have the maximal influence. For this reason, we first define the maximum influence perturbation set $\mathcal{P}_\mathbb{F}$ of a given subset of features $\mathbb{F}\subseteq\mathcal{F}$ as the set of perturbations that only perturbe features in $\mathbb{F}$ and have the maximal influence.  Further, the set of maximum influence perturbations is the union of $\mathcal{P}_\mathbb{F}$ for all feature sets:
{\small
  \begin{align*}
    \mathcal{P}_{\mathbb{F}}(d) &= \argmax_{p\in\mathbb{R}^n} I(d, p)\ s.t.\ \forall_{f_i \notin \mathbb{F}} \left(p_i = 0\right) \\
    \mathcal{P}(d) &= \bigcup_{\mathbb{F}\subseteq\mathcal{F}} \mathcal{P}_\mathbb{F}(d) 
  \end{align*}
}%
Based on these definitions, the total responsibility of a given feature $f_i\in\mathcal{F}$ is based on the responsibility of each perturbation that involves the feature. To this end, we define the responsibility $S^d_{f_i}$ of a feature $f_i$ for input point $d$ as the sum of all maximum influence perturbations that involve $f_i$ (e.g., the perturbation $p_i \ne 0$): 
{\small
  $$S^d_{f_i} = \sum_{p \in \mathcal{P}(d), p_i \ne 0} I(d, p)$$ 
}%
Putting this together, we can define the responsibility score $S^d_e$ of a feature explanation function (FEF) $e$ as the average of its bound features; where $\mathbb{F}_{e_i} \subseteq \mathcal{F}$ is the set of features bound to an FEF:
{\small
  $$S^d_{e} = \frac{\sum_{f_i\in\mathbb{F}_e} S^d_{f_i}}{|\mathbb{F}_{e}|}$$
}

We are now ready to present the key technical problem for text acquisition feedback:
\begin{problem}[Prescriptive Explanation]\label{p:problem}
	Given the feature vector of a data point $d$, prediction model $M$, a set of FEFs $E=\{e_1,\cdots,e_m\}$,  return the top $k$ FEFs whose responsibility is above a threshold $t$: 
{\small
	$$
  E^* = \topk_{e \in E}  S^d_e\ \ s.t.\ S^d_e > t
	$$
}
\end{problem}

\subsection{The {\large \sol} Heuristic Solution}

The space of solutions for Problem~\ref{p:problem} relies on enumerating all possible elements in the power set of the feature set $\mathcal{F}$, which is exponential in size: $2^{|\mathcal{F}|}$.  This  means that for $n$ features there are $2^{n}$ possible sets of (maximal influence) perturbations to naively explore. 

We instead present a heuristic solution called \sol whose complexity is linear in the number of paths in the random forest model.  The key insight is to take advantage of the structure of the random forest model to constrain the types of perturbations and feature subsets to consider.  A path is the sequence of decisions from the root of a tree to a leaf node.

The main idea is to scan each tree in the random forest and compute responsibility scores local to the tree.  
In addition, rather than compute the impact for all possible perturbations, we only consider the {\it minimal} perturbation with respect to each path in the tree.

Let $D=\{d_1,\dots,d_m\}$ be the training dataset and $Y=\{y_1,\dots,y_m\}$ be their labels.
The random forest model $M=\{T_1,\dots,T_t\}$ is composed of a set of trees.
A tree $T_i$ is composed of a set of $k$ decision paths $q_i^1,\dots,q_i^k$; each path $q_i^j$ matches a subset of the training dataset $D_i^j\subseteq D$ and its vote $v_i^j$ is the majority label in $D_i^j$.
Thus, the output of $T_i(d)$ is the vote $v_i^j$ of the path that matches $d$ (e.g., $d\in D_i^j$), and the output of the random forest $M(d) = \argmax_v |\{ 1 | v_i^j = v \}|$ is the majority vote of its trees.

Let $minp(d, q_i^j)$ return the minimum perturbation $p$ (based on its L2 norm) such that $d$ matches path $q_i^j$.  
{\small
  \begin{align*}
    minp(d, q_i^j) &= \argmin_{p \in \mathbb{R}^n} |p|_2\ s.t.\ q_i^j\ \textrm{matches}\ d+p
  \end{align*}
}%
Rather than examining all possible perturbations, our heuristic to compute $S^d_{f_i}$ restricts the set of perturbations with respect to the decision paths in the trees that increase $d$'s utility.  The impact function $I()$ is identical, however it takes a path $q_i^j$ as input and internally computes the minimum perturbation $minp(d,q_i^j)$.  This can be directly computed by examining the decision points along the path.  The confidence $C(d)$ is the fraction of samples in $D_i^j$ whose labels $y_k$ match the path's prediction $v_i^j$.    
{\small
\begin{align*}
  I(d, q_i^j)   &= \frac{U(v_i^j) - U(M(d))}{\Delta(minp(d, q_i^j))}\times C(d+minp(d, q_i^j))\\
  C(d) &= \frac{|\{d_k \in D_i^j | y_k = v_i^j \}|}{|D_i^j|}
\end{align*}
}%
If two paths within a tree perturb the same set of features, we only consider the path with the maximal impact score.  In addition, we do not compare paths across trees. 
We define $Q_i$ as the set of {\it maximal impact paths} of a tree $T_i$, with at most one path for a given subset of features.
$\mathbb{F}_p$ is the subset of features that $p$ perturbs:
{\small
\begin{align*}
  Q_i(d) &= \{ q \in T_i |\forall_{q' \in T_i} I(d,q) \ge I(d,q') \wedge \mathbb{F}_{minp(d,q)} = \mathbb{F}_{minp(d,q')} \}\\
  \mathbb{F}_p &= \{ f_i \in \mathcal{F} | p_i \ne 0  \}
\end{align*}
}%
Finally, $S^d_{f_i}$ computes the responsibility for $f_i$ as the sum of all maximal influence paths in all decision trees that improve the predicted utility $U()$.
{\small
\begin{align*}
  S^d_{f_i} &= \sum_{T_i \in M} \sum_{q_i^j \in Q_i(d)} I(d, q_i^j)  \ \textrm{if}\  U(v_i^j) > U(M(d))
\end{align*}
}%
Our implementation indexes all paths in the random forest by their utility.  
Given $d$ and predicted utility $U(M(d))$, we retrieve and scan the paths with higher utility.
For each scanned path $q$, we compute the change in the utility function, discount its value by the minimum perturbation $p$ as well as the path's confidence.  
We then select the maximal impact paths for each tree; for each path, we add the responsibility score of all features perturbed in its minimum perturbation~$p$.  The final scores are used to select from the library of explanation functions.  

\stitle{Normalization:}  We find that features closer to the root will happen to occur in more feature sets and have artificially higher scores, thus we need to adjust feature impact scores to reduce bias.  To do so, we draw a sample of text from the corpus that has been labeled as low quality.  For each feature $f_i$, we compute the responsibility for each low quality text, and aggregate their values to compute the sample mean $\mu_{f_i}$ and standard deviation $\sigma_{f_i}$.  We then normalize a feature's responsibility $S^d_{f_i}$ by computing $Snorm^d_{f_i} = \frac{S^d_{f_i} - \mu_{f_i}}{\sigma_{f_i}}$. 

\stitle{Picking FEFs:} Once the feature scores have been computed, identifying the top-k FEFs is straightforward, and we compute each FEF's average impact score using a series of fast matrix operations. Let $\vec{s}\in \mathbb{R}^{n}$ where $\vec{s}_{i}=Snorm^d_i$, and matrix $A\in\mathbb{R}^{m \times n}$ represent the features bound to each of the $m$ FEFs, where $A_{ji}=1$ if feature $f_i$ is bound to FEF $e_j$, otherwise 0. Also, let $\vec{e}\in \mathbb{R}^{m} = A\vec{s}$. $\frac{\vec{e}_j}{\sum_{i=1}^{n} A_{ji}}$ is the average impact score of all features mapped to the $j$th FEF. We then sort the FEFs by their average scores and take the top k with a score above the threshold $t$.  

\section{New Application Domains}\label{s:extend}

How much work does it take to add rich feedback support for text in a new domain?  We describe our process to extend \sys to two domains with different quality measures: product reviews that care about helpfulness to a shopper~\cite{archak2011deriving}, and then host profiles that are judged by trustworthiness to renters~\cite{ma2017self}.    
We start with the feature library of $47$ features and no explanation functions.

The general approach is to survey quality assessment research in a domain to borrow useful features and explanations.
We did not require new features for product reviews; we simply label reviews with $\geq 60\%$ helpful votes as high quality and low otherwise.   The resulting model ($85\%$ accuracy, balanced test set) was competitive with existing work~\cite{ghose2007designing}. 

For explanation functions, prior work showed that $75\%$ of reasons for unhelpful reviews were covered by (in priority order) overly emotional/biased opinions, lack of information/not enough detail, irrelevant comments, and poor writing style~\cite{badreviewreasons2}. These naturally map to 4 of our feature categories, so we wrote explanation functions for each and bound them to the features in the corresponding category.   For instance, the following defines the function for Off-Topic text:
{\small
 \begin{verbatim}
def offTopic(topics, text="", feats=[]):
  if len(topics) < 5:
    sortedTopics = sorted(topics, key=topic.prob)
    return "Try discussing some of these topics: " +
           topK(sortedTopics, 5)
\end{verbatim}}
We used a similar process for host profiles and found that research emphasizes trustworthiness as the key quality metric~\cite{ma2017icwsm, ma2017self}.  Their work identified a subset of the Linguistic Inquiry and Word Count (LIWC) features~\cite{pennebaker2015development} and other features as useful for measuring trustworthiness.    The primary groups of features related to absence of detail and low topic diversity.  Reading through their table of features, we also found that writing style and friendliness features were common.  

We added LIWC API calls to \sys; the model tested on a balanced set of $300$ AirBnB host profiles was competitive ($79\%$ accuracy) at predicting if a profile was $\geq$ median trustworthiness.  All trustworthiness factors except friendliness directly corresponded to existing explanation functions.  Thus, we wrote a friendliness explanation function that suggested writing more friendly and inclusive prose, and bound it to the relevant LIWC features (social, inclusive, etc).

Thus, three of the FEFs, Informativeness, Topic, and Readability/Grammar, overlapped between the two domains. The fourth FEF for product reviews was mapped to Subjectivity features in (Table 1) and the fourth host profiles FEF was mapped to Friendliness LIWC features shown in~\cite{ma2017icwsm}, with each returning text suggesting that the user improves the respective facet of their submission (i.e., {\it ``Please make your writing more balanced and neutral''}). Other explanation functions (Topic, Informativeness) suggested specific content for the user to write about, mined from high-quality documents from each corpus (i.e., topics, jargon). 

Overall, each explanation function was 3-20 lines of python code.  We are optimistic about the \isep pattern, because adopting to new domains is simply a matter synthesizing existing research by adding features and creating simple explanation functions.

\section{Experiments}
\label{s:exp}

We now evaluate how \sys improves high-quality data acquisition using live Mechanical Turk deployments.  First, we validate the value of \prehoc by running a crowdsourced data acquisition experiment with different \qp optimizations for foreign key and domain constraints.  Second, we evaluate \sys's \isep pattern for text acquisition in two domains---acquiring customer reviews for Amazon products~\cite{amazondataset} and acquiring profile descriptions for AirBnB host profiles~\cite{ma2017icwsm}.  \sys is able to adopt to the domains' different quality measures (helpfulness vs trustworthiness) with small configuration changes. Finally, we perform a detailed analysis of how segmentation and \sol each contribute to improving the quality of the acquired text.

\subsection{Precog for Hard Constraints}

\begin{figure}[h]
\centering
\includegraphics[width=.7\columnwidth]{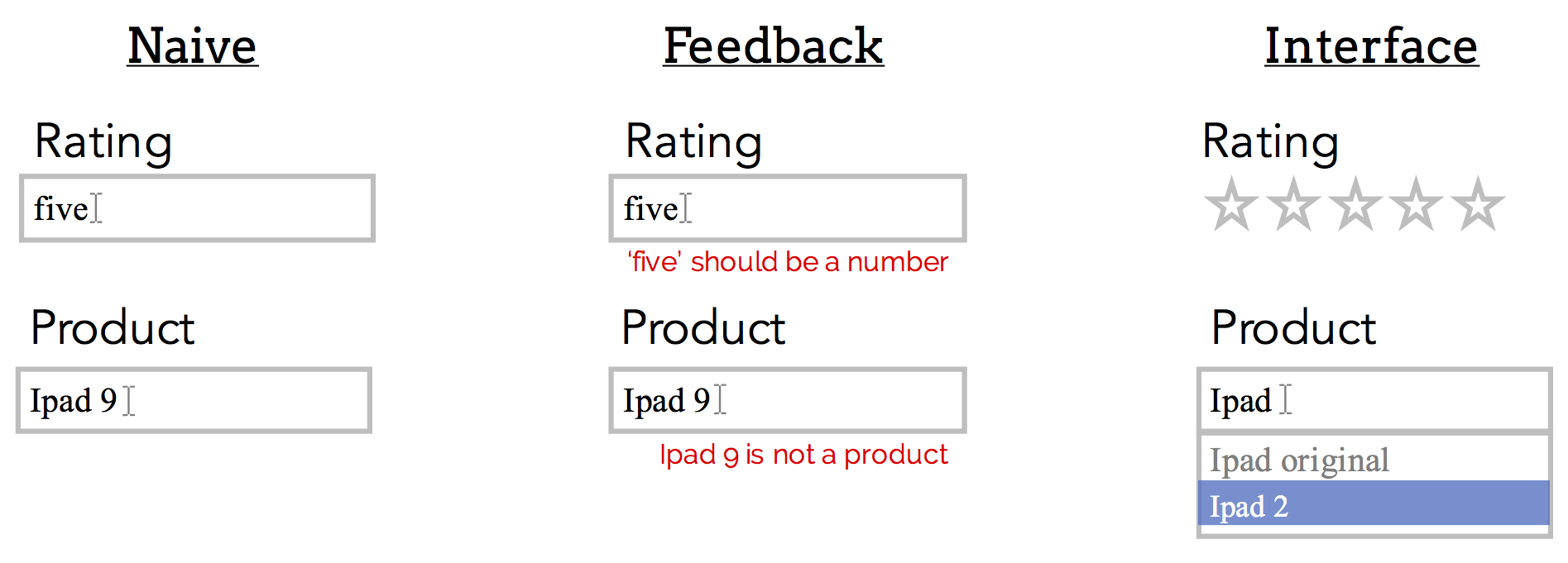}
\caption{\small Worker interfaces to evaluate no optimization, custom feedback \qp, and custom interface \qp for hard constraints.}
\label{f:exp_qpd_interface}
\end{figure}

Although it is intuitively obvious that form feedback and custom interfaces should improve quality, we quantify the amount using the example from Section~\ref{s:arch}.
We evaluate \qp for \texttt{product\_id} (foreign key constraint) and \texttt{rating} (domain constraint) from the \texttt{reviews} table.  Figure~\ref{f:exp_qpd_interface} depicts the three interfaces that are created---naive with no \qp, customized feedback, and customized interface optimizations.  

We created a simple Mechanical Turk task that asked workers to submit the product model of their cell phone along with a 1 to 5 rating for the phone's quality; each worker was paid $\$0.05$ to complete the task. Each worker was randomly assigned to one of three conditions, one for each of the interfaces shown in Figure~\ref{f:exp_qpd_interface}. The experiment was run until $100$ workers had participated in each condition. For the foreign key constraint, we populated a \texttt{products} table with all cell phone product models from the Amazon product corpus and a comprehensive list of phone models~\cite{nfc}. We relaxed the foreign key constraint by ignoring case sensitivity of the product names.

\begin{figure}[h]
\centering
\includegraphics[width=.6\columnwidth]{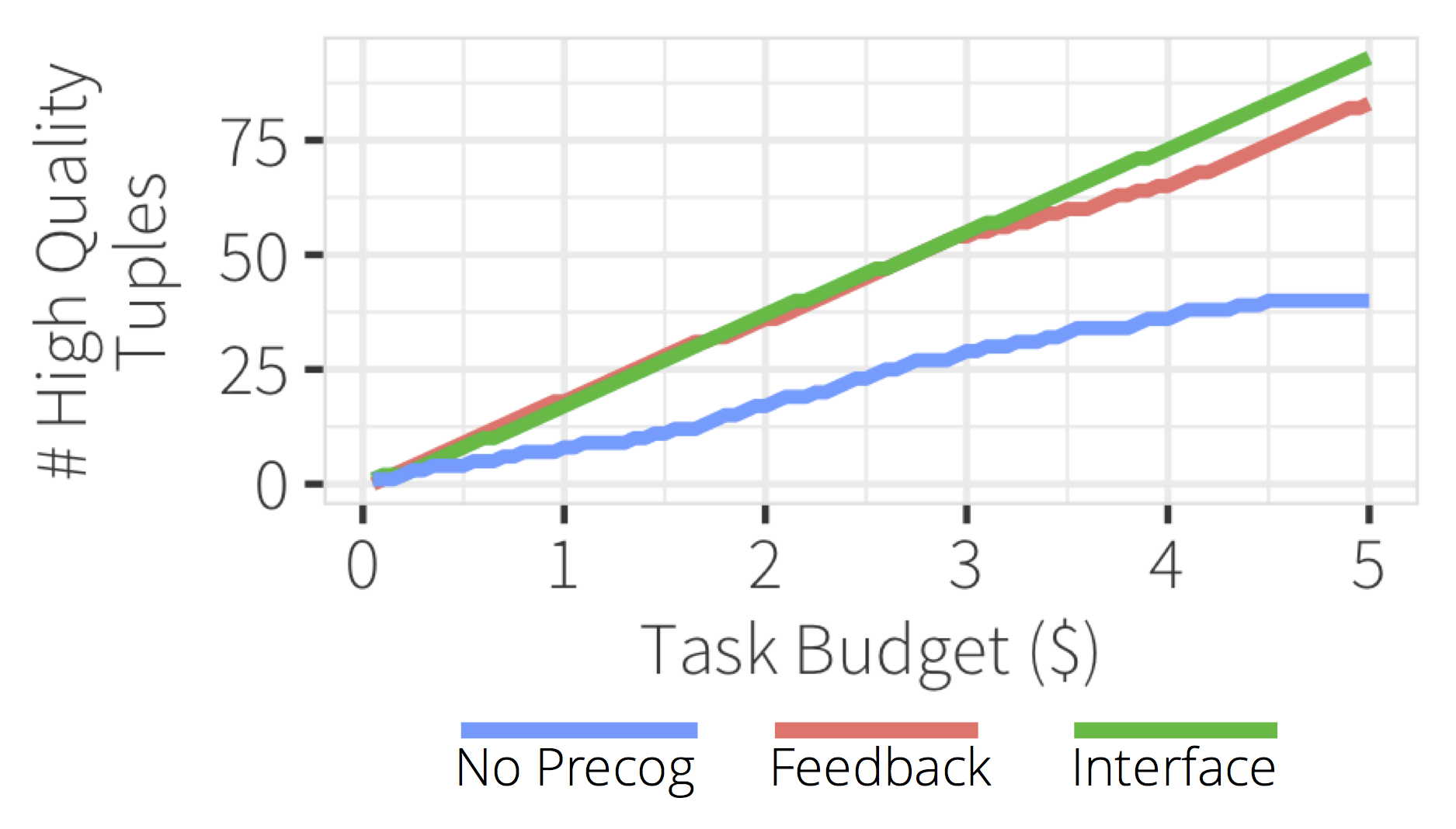}
\caption{\small \# records satisfying both constraints vs budget.
Feedback and interface customization acquire $1.7\times$ and $1.9\times$ more valid records than no \qp optimization.}
\label{f:exp_qpd_cost_cdf}
\end{figure}

Figure~\ref{f:exp_qpd_cost_cdf} plots the number of high quality tuples that were collected as a function of the number of completed tasks; we define a tuple as high quality if no constraints were violated.  Feedback and interface customization acquire $1.7\times$ and $1.9\times$ more high quality tuples than no \qp optimization.

\subsection{Precog for Text Acquisition}

\stitle{Setup and Datasets:} \sys is setup as described in Section~\ref{s:extend}: we train \sys using the laptop category of the Amazon product reviews corpus~\cite{amazondataset}, and the AirBnb profile corpus~\cite{ma2017icwsm} for their corresponding experiments. We then synthesized existing research to write 4 explanation functions for each domain, with 3 overlapping between the two.

\stitle{Procedures:} 
Participants writing product reviews were asked to write a {\it review of their most recently owned laptop computer} ``as if they are trying to help someone else decide to buy that laptop or not and are writing on a review website like the Amazon store''. We used a qualification task to ensure participants had ever owned a laptop. Participants writing Airbnb profiles were asked to ``pretend that [they] are interesting in being a host on Airbnb'' and to ``write an Airbnb profile for [themselves]''. Participants were told that upon submitting their writing, they {\it may} receive feedback and could {\it optionally} revise.

Upon pressing the \texttt{I'm Done Writing} button, the interface displayed our document-level feedback under the text field; for users in the segmentation condition, low quality segments were highlighted red and the related feedback displayed when users hovered over the segment. We then gave participants the opportunity to revise their submission; to avoid bias, we noted that they were ${\textbf{not}}$ obligated to.   At this point, users could click the \texttt{Recompute Text Feedback} button (median 1 click/participant), or press \texttt{Submit} to submit and finish the task. We used a post-study survey to collect demographic information as well as their subjective experience.

The interface was the same for all conditions---only the feedback content changed.  The final  submission was considered the {\it post-feedback} submission, and the initial submission upon pressing the \texttt{I'm Done Writing} was the {\it pre-feedback} submission.  The experiment was IRB approved.

\stitle{Experimental Conditions:}
The purpose of experiments is to both show the Cost Saving benefits of \sys as well as to evaluate the effectiveness of it's two main features (segment-level feedback and \sol explanation generation). We thus assign each participant to one of four conditions. A detailed explanation of the four conditions is shown in Section 7.2.2. We first present the results of the fully featured \sys condition (Section 7.2.1) and then demonstrate the contribution of each \sys component (in Section 7.2.2).

\stitle{Product-Review Participants:}
For the laptop review experiment, we recruited $85$ workers on Amazon's Mechanical Turk (61.2\% male, 38.8\% female, ages 20-65 $\mu_{age}$=32, $\sigma_{age}$=8.5). $81$ completed the task. Participants were randomly assigned to one condition group; all conditions had 21 subjects except the \emph{\sys} condition which had 22.  No participant had used \sys before. $71.3\%$ had written a prior product review; all had read a product review in the past. All participants were US Residents with > 90\% HIT accept rates. The average task completion time was 14 minutes, and payment was $\$2.5$ ($\sim\$10/hr$).

\stitle{Host-Profile Participants:}
For the profile description experiment, we recruited $92$ workers on Amazon's Mechanical Turk (58.7\% male, 41.3\% female, ages 20-62 $\mu_{age}$=33, $\sigma_{age}$=8.2); all completed the task. Participants were randomly assigned to one condition group; with (21,26,22,23) participants in conditions (1,2,3,4), respectively.  No participant had used \sys before. $62\%$ had used AirBnb before. All participants were US Residents with > 90\% HIT accept rates. The average task completion time was 11 minutes, and payment was $\$2.5$ ($\sim\$13.6/hr$).

\stitle{Protocol and Rubric for Assessing Quality:}
Three independent evaluators (non-authors) coded the pre {\it and} post-feedback documents using a rubric based on prior work on review quality~\cite{badreviewreasons2,mudambi2010makes,liu2007low} and Airbnb profile quality ~\cite{ma2017self}. Each rubric rated documents on a 1-7 Likert scale using three specific aspects identified by prior work---Informativity, Subjectivity, Readability---for reviews---Ability, Benevolence, Integrity--for profile trustworthiness, as well as a holistic overall score. The change in these measures between pre and post-feedback suggests the utility of the feedback.

The review rubric asks coders to scores reviews on helpfulness to laptop shoppers, and the host profile rubric asks coders to score profiles based on trustworthiness to potential tenants. Each defines the three main measures, and provides examples that contribute positively and negatively to each criteria. 

For product reviews, {\it Informativity} is the extent that the review provides detailed information about the product, where 7 means that the review elaborates on all or almost all of the specifications of a product while 1 means that it states an opinion  but fails to provide factual details (e.g., laptop specifications). {\it Subjectivity} is the extent that the review is fair and balanced but with enough helpful opinions for the buyer to make an informed decision: 1 means the review is an angry rant or lacks any opinions while 7 means it is a fair and balanced opinion. {\it Readability} is the extent that the review facilitates or obfuscates the writer's meaning.  For instance, a review that consists of many ambiguous phrases like ``I have never done anything crazy with it and it still works.'' is assigned 1 as it might require multiple readings to understand. {\it Overall Quality} is the holistic helpfulness of the review for prospective buyers. 

Ma et. al describe the meaning of the three Airbnb criteria in ~\cite{ma2017self}: {\it Ability} ``refers to the host's domain specific skills or competence.'' {\it Benevolence} ``refers to the host's domain specific skills or competence.'' {\it Ability} ``refers to the host's domain specific skills or competence.'' Each measure is rated on a scale from 1 (Strongly Disagree) to 7 (Strongly Agree) based on coder agreement with a set of statements mapped to each criterion (i.e., ``This person will stick to his/her word, and be there when I arrive instead of standing me up'' for integrity). The full set of coder statements is described at length in ~\cite{ma2017self}. {\it Overall Quality} is the holistic trustworthiness of the host for prospective tenants. 

Finally, we asked coders to subjectively rate their agreement from 1-7 to the statement  {\it``The post-feedback revisions improved on the pre-feedback document.''}, or 0 if the document did not change. Each measure is the average of the ratings from two coders---if they differed by $\geq 3$, a third expert coder was used as the tie breaker and decided the final value.  The third coder was trained by being shown the Amazon or Airbnb corpus, examples across the quality spectrum, and the other two coders.  The coders labeled documents in random order and did not have access to any other information about the documents.

\subsubsection{Cost Savings}

\begin{figure}[th]
\centering
\includegraphics[width=\columnwidth]{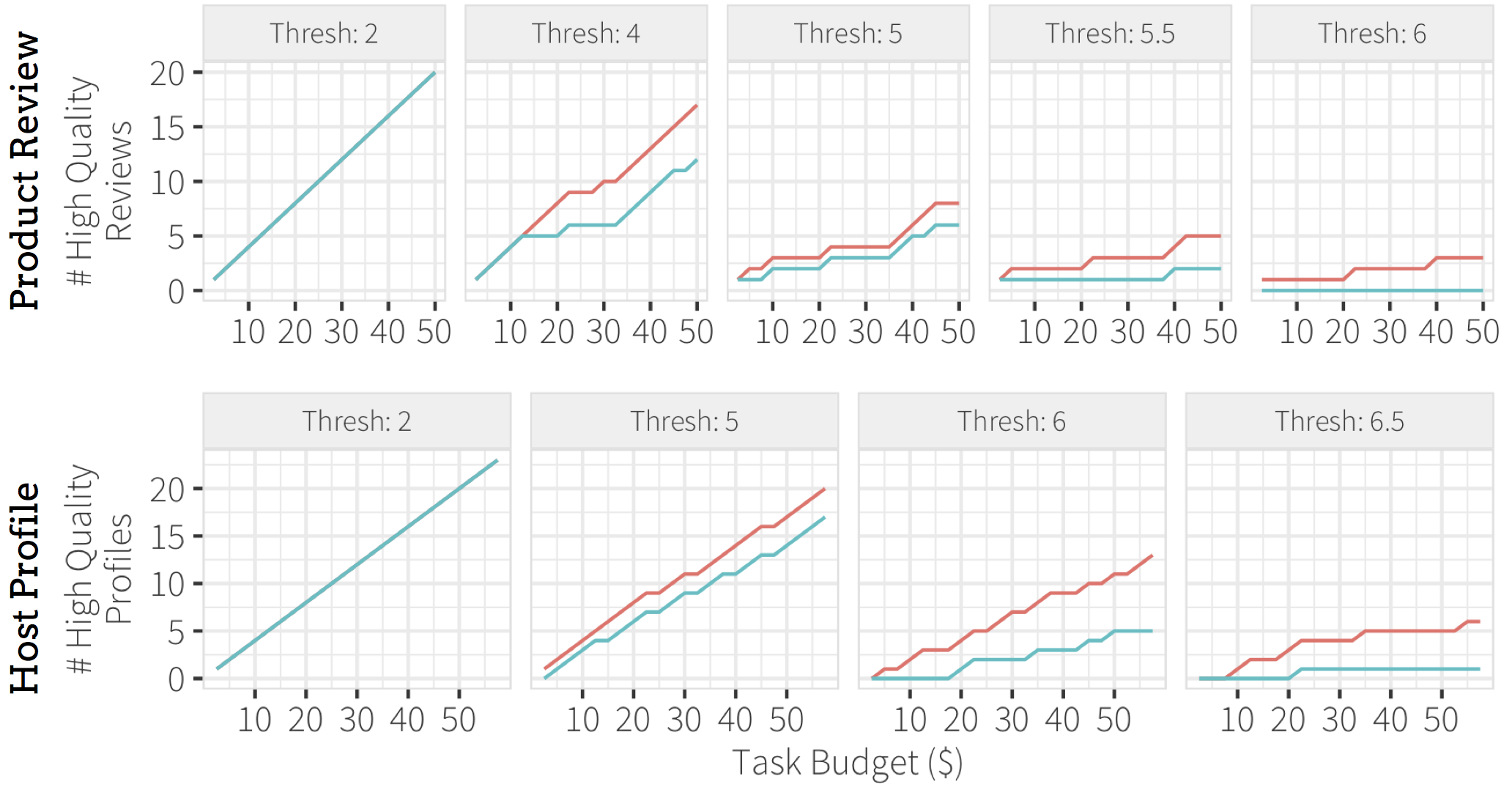}
\caption{\small \# of documents where $quality \ge thresh$, for varying thresholds; product reviews (top), host profiles (bottom). 
\qp is more effective than no \qp when the desired quality is high.}
\label{f:exp_both_cost_cdf}
\end{figure}

Figure~\ref{f:exp_both_cost_cdf} compares \qp against the baseline of not using \qp (naive review collection). 
We plot CDF curves for the number of high quality documents as the task budget increases.
Each facet defines high quality at a different threshold; product reviews and host profiles are shown as the top and bottom rows, respectively.  
When the threshold is low, it is easy to acquire low-quality text and both approaches are the same.
However, \qp is more effective when the threshold increases.
For the reviews and profiles experiments, \qp acquires $\ge2\times$ and $\ge2.6\times$ more high quality documents than the baseline for thresholds of $5.5$ and $6$, respectively.
Note that the baseline does not acquire {\it any} high quality reviews when $\textrm{thresh}\ge6$. \sys only marginally increases latency of each worker. The average host profile took $6.8$ minutes to complete without \sys, and $11.1$ minutes with the additional feedback from \sys. Similarly, Airbnb profiles took an average of $10.2$ minutes to complete without \sys and $15.3$ minutes to complete with \sys. Such latency difference is relatively small if we compare the end-to-end time of two systems since the majority of the time was spent on worker recruitment.

\subsubsection{Segment, Explain, or Both?}

Are both Segment and Explain necessary in the \isep pattern?
To understand the contributing factors towards the quality improvements, we compared four feedback systems that varied along two dimensions: {\it granularity} varies the feedback to be at the document level (\texttt{Doc}), or at the document {\it and} segment level (\texttt{Seg}); {\it explanation selection} compares the single-feature outlier technique from~\cite{krause2015method} (\texttt{Krause}) with \sol.  This results in a 2x2 between-subjects design. {\it \sys} denotes the segment-level \sol-based system.

\texttt{Krause}~\cite{krause2015method} was shown to outperform static explanations of important components of a helpful review (similar to a rubric) for students performing peer code-reviews and uses an outlier based approach described in Section~\ref{s:explain-related}.  To ensure fair comparison, we supplemented their features with domain-specific features for Informativeness (\# of product features/jargon), Readability (Coleman-Liau index), and Friendliness (LIWC features related to friendliness) so that their features are comparable to those used in our feature library.

To summarize, each participant was randomly assigned to one of four conditions: \emph{Doc+Krause}, \emph{Seg+Krause}, \emph{Doc+\sol} and \emph{\sys} (\emph{Seg+\sol}).

\begin{figure}[h]
\centering
\includegraphics[width=\columnwidth]{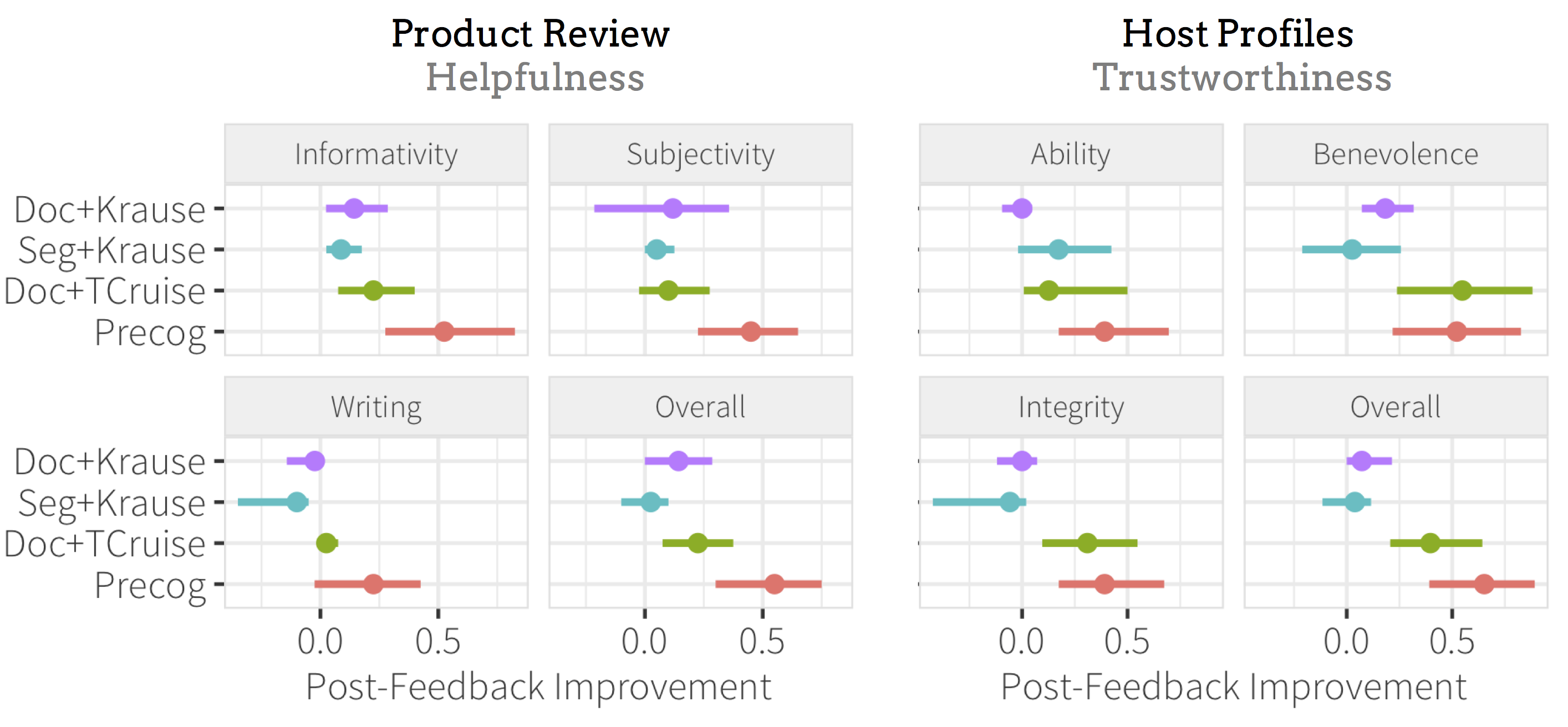}
\caption{\small Improvement on Likert scores for both domains (reviews and profiles) and four quality criteria per domain. Note that the quality criteria differ across domains.}
\label{f:improvement}
\end{figure}

\begin{figure}[h]
\centering
\includegraphics[width=\columnwidth]{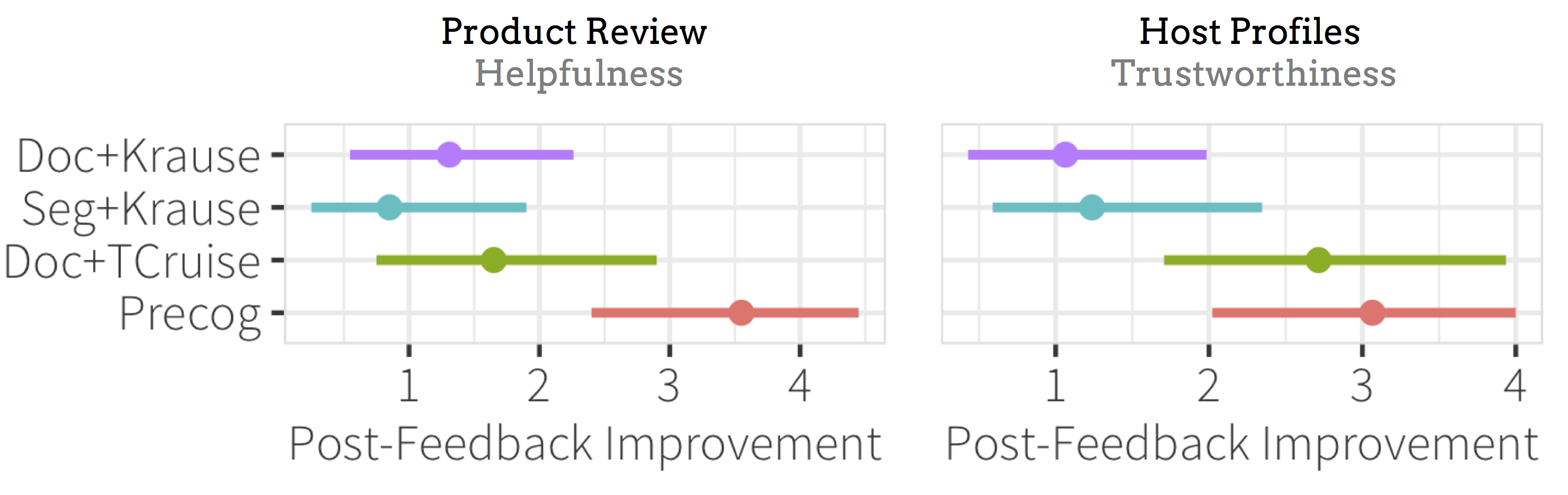}
\caption{\small Subjective agreement to: ``The post-feedback revisions improved on the pre-feedback review.'' for product reviews, and ``The post-feedback revisions are more trustworthy than the pre-feedback profile.'' for host profiles.}
\label{f:improve2}
\end{figure}

Figure~\ref{f:improvement} plots the mean change and $95\%$ boostrap confidence interval for the four rubric scores. Figure~\ref{f:improve2} shows a similar chart for the coder's subjective opinion of the improvement. These plots show the effect size across all measures, and that the largest improvements were due to the combination of segmentation {\it and} \sol-based explanation.

We conduct statistical tests to further investigate the results. For both Product Review and Host Profiles, we performed Two-Way ANOVAs with both the Overall Quality Improvement and Subjective Coder Improvement Scores as the dependent variables, and \sol and segmentation as the independent variables. We then performed pairwise Tukey HSD post-hoc tests between each of the four conditions.

We found that  combining segmentation {\it and} \sol-based explanation outperformed all other conditions by a statistically significant margin for Product Reviews, and outperformed all but the next-best {\it Doc+\sol} condition for Host Profiles. Furthermore, controlling for the other variable, \sol showed a statistically significant difference in improvement, while segmentation did not.

However, the combination of segmentation and \sol consistently produced larger effect sizes than all other conditions across both Host Profiles and Product Reviews: for Product Reviews \sys, which combines segmentation and \sol, improved the overall measure (bottom left facet) by nearly $3.9\times$ over the baseline (0.55 vs. 0.14 increase), and a $2.4\times$ improvement over the next-best {\it Doc+\sol} condition. For Host Profiles \sys improved the overall measure by nearly $7.1\times$ over the baseline (0.65 vs. 0.07 increase), and a $1.7\times$ improvement over the next-best {\it Doc+\sol} condition. 

In summary, we find that \sol is essential to improving document quality; combining \sol with Segmentation empirically produces the best results across the board.

\section{Related Work}
\label{s:related}

Sections~\ref{s:predict} and~\ref{s:explain} surveyed work related to text quality prediction and writing feedback.  We now describe related work in terms of data acquisition interface optimizations, quality control in crowdsourcing and other post-hoc quality mechanisms specific for text acquisition.

\stitle{Survey Design and Optimization: } 
The survey design literature has studied ways to re-ordering, and designing survey forms in order to reduce data entry errors.  These include guidelines and constraints on form elements~\cite{groves2011survey,norman2003online}, as well as interface techniques such as double entry~\cite{day1998double} commonly used for picking passwords.  These can be integrated as feedback and interface customizations in \sys.
  
A closely related work from the database community is Usher~\cite{chen2011quality}, which have similar goals to improve data collection quality.  Usher analyzes an existing corpus of collected data to dynamically learn soft constraints on data values, and focuses on input placement, re-asking, and some interface enhancements.  These ideas can be viewed as instances of \qp.  
To contrast, we focus on using explicit constraints and ambiguous quality measures (for text) and provide explicit \texttt{DDL} statements to push them to the input interface.  Additionally, our \isep pattern addresses on free-form text entry that complements their focus on simple data types.

\stitle{Quality Control in Crowdsourcing:} Quality control is an important research topic in crowdsourced data management~\cite{DBLP:journals/tkde/LiWZF16,DBLP:journals/tkde/ChittilappillyC16,DBLP:journals/tkde/Garcia-MolinaJM16}. It has been extensively studied in recent years~\cite{DBLP:conf/sigmod/SarmaPW16,DBLP:conf/sigmod/FanLOTF15,DBLP:conf/sigmod/ZhengWLCF15,DBLP:journals/pvldb/CaoSTC12,DBLP:conf/icde/BoimGMNPT12,DBLP:journals/pvldb/GaoLZFH15,haas2015argonaut}. There are some works that apply \prehoc to improving crowd quality~\cite{whang2012compare,DBLP:conf/icde/TrushkowskyKFS13,DBLP:conf/sigmod/ParkW14}.  Further, review hierarchies were proposed for hierarchical crowdsourced quality control using expert crowds~\cite{haas2015argonaut}.   However this work either focuses on a particular application~\cite{whang2012compare}, or not intended to support custom interfaces~\cite{DBLP:conf/sigmod/ParkW14}.   Moreover, none focus on multi-paragraph text attributes such as product reviews or forum comments. To the best of our knowledge, \sys is the first system that systematically supports \qp for a wide range of data types and quality specifications (constraints and quality scores).

\stitle{Post-hoc Approaches for Text Acquisition: }  A dominant approach is to filter poor content~\cite{spirin2012survey} such as spam; sort and surface higher quality content~\cite{tang2013social,guy2015social,agichtein2008finding} such as product reviews~\cite{mudambi2010makes}, answers to user comments~\cite{wang2013wisdom,yang2016recommending}, or forum comments~\cite{siersdorfer2010useful}; or edit user reviews for clarification or grammatical purposes~\cite{bernstein2010soylent,kittur2011crowdforge,zapposedit}.  These approaches incur additional quality control costs and are complementary to \qp.  They also assume a large corpus that contains high quality content for every topic (e.g., product or question). In reality, there is often a long tail of topics without sufficient content for such approaches to be effective~\cite{quoraanswers,amazondataset}. For such cases, improving quality during user input process may be more effective.

\stitle{Indirect Quality Mechanisms: }
Indirect methods such as community standards and guidelines~\cite{nov2007motivates,bakshy2009social,amazonguidelines} help clarify quality standards, while up-votes and ratings provide social incentives~\cite{muchnik2013social,bosu2013building}.    Incentive mechanisms such as badges, scores~\cite{ghosh2012social,deterding2011game}, status~\cite{zappospremium}, or even money~\cite{kim2012institutionalization,zapposedit} have also been used to keep good contributors.  These methods focus more on finding good contributors and lack content-specific feedback (e.g., discuss camera quality for a phone).

\section{Conclusion and Future Work}
\label{s:conclusion}

This paper presented the design, implementation and evaluation of  \sys, a pre-hoc quality control system. The basic idea is to push data-quality constraints down to the data collection interface and improve data quality before acquisition. While the idea is easy to achieve for simple data types and constraints, it faces significant challenges for text documents. We address these challenges by proposing a novel segment-predict-explain pattern for detecting low-quality text and generating {\it prescriptive} explanations to help the user improve their text. Specifically, we develop effective approaches to measure text quality at both document and segment levels, present an efficient technique to solve the prescriptive explanation problem, and discuss how to extend \sys to new domains. Through extensive MTurk experiments, we find that \sys collects $\geq 2 \times$ more high-quality documents and improves text quality by 14.3\% compared to not using pre-hoc techniques.

Though \sys demonstrates the feasibility of such automated interfaces, it also reveals several areas of improvement.  Due to a small number of explanation functions, study participants found that repeatedly using the system began to provide redundant feedback; simplifying the development of more explanation functions may help the system produce more nuanced feedback. We also used document-quality labels to train the segment classifier. We showed this to be sufficient by testing on crowd-sourced labels; however more sophisticated techniques to classify segments could improve feedback. 


In the long term, we envision \sys as an example of automatically applying \prehoc (e.g., writing feedback) based on downstream application needs (e.g., quality reviews). 
In future work, we hope to explore a broader range of applications (e.g., different social media domains or user contexts), and study how to optimize data-collection interfaces to meet more complex application needs.

\balance
{
\small
\bibliographystyle{abbrv}
\bibliography{main}
}

\end{document}